\documentclass[aps,prl, notitlepage,reprint,superscriptaddress]{revtex4-1}

\usepackage{changes}

\usepackage[colorlinks,linkcolor=red,citecolor=blue,urlcolor=red]{hyperref}
\renewcommand{\t}[1]{\mathrm{#1}}

\usepackage{graphicx}
\usepackage{ulem}
\usepackage{amsmath,amssymb,amsfonts} 
\usepackage{bm}	
\renewcommand{\mathbf}{\bm}

\usepackage{dsfont}	
\renewcommand{\mathbb}{\mathds}	

\usepackage{mathrsfs} 
\usepackage{mathtools} 

\usepackage{color}

\newcommand{\fref}[1]{Fig.~\ref{#1}}
\renewcommand{\eqref}[1]{Eq.~\ref{#1}}

\begin{document}
	
	\title{Integrated gallium phosphide nonlinear photonics}
	\author{Dalziel J. Wilson}
	\thanks{These authors contributed equally to this work}
	\affiliation{IBM Research --- Zurich, Sa\"{u}merstrasse 4, 8803 R\"{u}schlikon, Switzerland}
	\affiliation{Institute of Physics (IPHYS), {\'E}cole Polytechnique F{\'e}d{\'e}rale de Lausanne, 1015
		Lausanne, Switzerland}
	
	\author{Katharina Schneider}
	\thanks{These authors contributed equally to this work}
	\affiliation{IBM Research --- Zurich, Sa\"{u}merstrasse 4, 8803 R\"{u}schlikon, Switzerland}
	
	\author{Simon H\"{o}nl}
	\thanks{These authors contributed equally to this work}
	\affiliation{IBM Research --- Zurich, Sa\"{u}merstrasse 4, 8803 R\"{u}schlikon, Switzerland}
	
	\author{Miles Anderson}
	\affiliation{Institute of Physics (IPHYS), {\'E}cole Polytechnique F{\'e}d{\'e}rale de Lausanne, 1015
		Lausanne, Switzerland}
	
	\author{Tobias J. Kippenberg}
	\affiliation{Institute of Physics (IPHYS), {\'E}cole Polytechnique F{\'e}d{\'e}rale de Lausanne, 1015
		Lausanne, Switzerland}
	
	\author{Paul Seidler}
	\email{pfs@zurich.ibm.com}
	\affiliation{IBM Research --- Zurich, Sa\"{u}merstrasse 4, 8803 R\"{u}schlikon, Switzerland}
	
	\date{\today}
	
		\begin{abstract}
			
			Gallium phosphide (GaP) is an indirect bandgap semiconductor used widely in solid-state lighting.  Despite numerous intriguing optical properties---including large $\chi^{(2)}$ and $\chi^{(3)}$ coefficients, a high refractive index ($>3$), and transparency from visible to long-infrared wavelengths ($0.55-11\,\mu$m)---its application as an integrated photonics material has been little studied.  Here we introduce GaP-on-insulator as a platform for nonlinear photonics, exploiting a direct wafer bonding approach to realize integrated waveguides with 1.2 dB/cm loss in the telecommunications C-band (on par with Si-on-insulator). High quality $(Q> 10^5)$, grating-coupled ring resonators are fabricated and studied.  Employing a modulation transfer approach, we obtain a direct experimental estimate of the nonlinear index of GaP at telecommunication wavelengths: $n_2=1.2(5)\times 10^{-17}\,\t{m}^2/\t{W}$. We also observe Kerr frequency comb generation in resonators with engineered dispersion. Parametric threshold powers as low as 3 mW are realized, followed by broadband ($>100$ nm) frequency combs with sub-THz spacing, frequency-doubled combs and, in a separate device, efficient Raman lasing. These results signal the emergence of \mbox{GaP-on-insulator as a \textcolor{black}{novel} platform for integrated nonlinear photonics.}
		\end{abstract}

\maketitle

\begin{figure}[t!]
	\includegraphics[width=1.0\columnwidth]{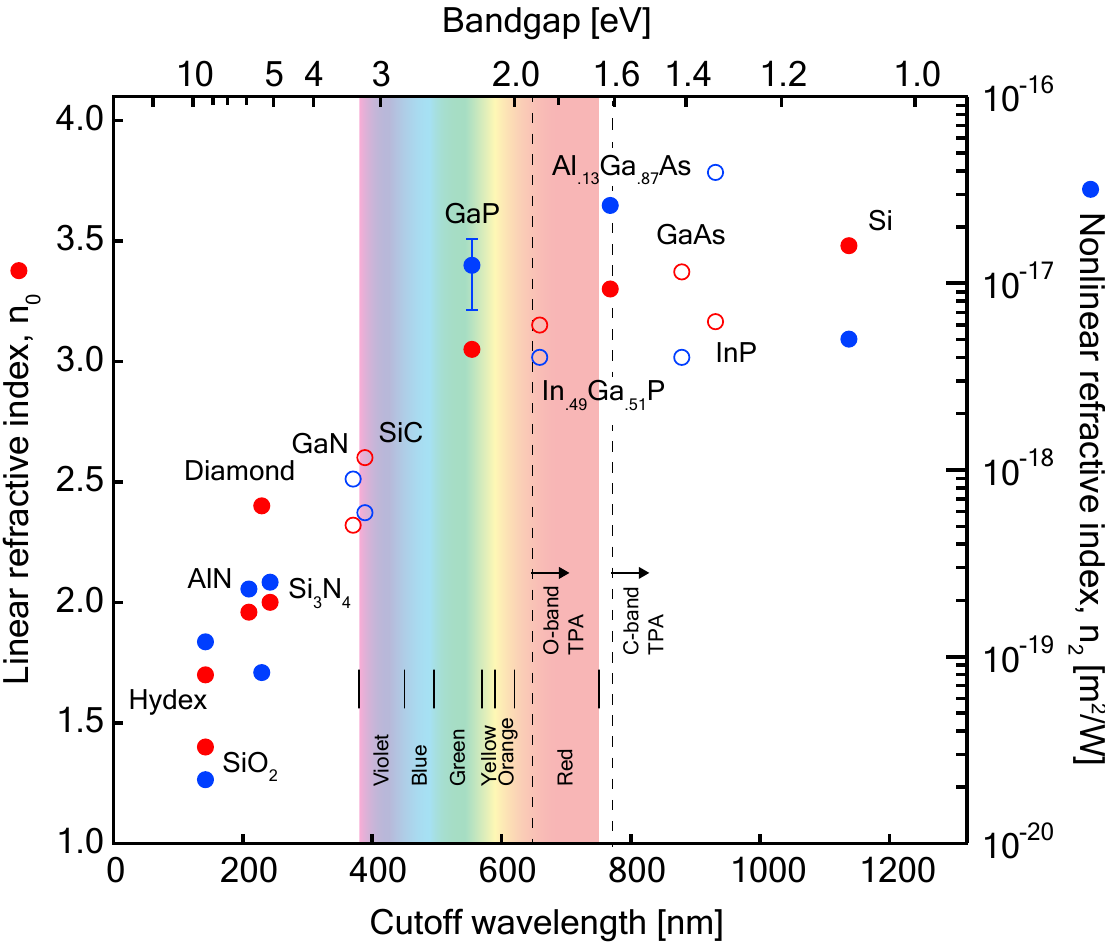}
	\caption{\color{black} \textbf{GaP as a material for integrated nonlinear photonics}. Linear (red) and nonlinear (blue) refractive indices of various integrated photonics materials, plotted versus bandgap energy and cutoff wavelength.  
		Vertical lines denote cutoff wavelengths for two-photon absorption in the O-band (1260-1360 nm) and C-band (1530-1565 nm). Solid circles denote materials used for microresonator frequency combs. Values for materials other than GaP were taken from \cite{pu2016efficient} (SiO$_2$, Hydex, AlN, Diamond, Si$_3$N$_4$, AlGaAs), \cite{lu2014optical} (SiC), \cite{dinu2003third} (Si), \cite{sun2000third} (GaN), \cite{ching1993calculation} (InP), \cite{martin2017gainp} (InGaP), and \cite{hurlbut2007multiphoton} (GaAs).  }
	\label{figure1}
	\vspace{-5mm}
\end{figure}

\small
Gallium phosphide has played an important role in the photonics industry since the 1960s, forming the basis for a range of light-emitting devices---most notably green LEDs---despite possessing an indirect bandgap \cite{pilkuhn1966green}.  More recently, efforts have been made to realize nanophotonic devices in GaP.  The motivation is severalfold: First, GaP is nearly lattice-matched to silicon, in principle enabling waferscale production \cite{mori1987new}. Second, GaP has negligible two-photon absorption (TPA) for wavelengths above 1.1 $\mu$m, enabling high power operation in all telecommunications bands. Third, among visibly transparent III-V materials, GaP has the largest refractive index ($n_0>3$), enabling strong optical confinement and implying a large $\chi^{(3)}$ nonlinearity (\fref{figure1}). Finally, the non-centrosymmetric crystal structure of GaP yields a nonzero piezo-electric effect and large $\chi^{(2)}$ nonlinearity.  Among the diverse applications \textcolor{black}{made possible} by these features, particular attention has been paid to realizing frequency doublers from telecommunication to visible wavelengths \cite{rivoire2009second,lake2016efficient}, nano-antennae for enhanced photoluminescence \cite{gan2012high}, and interfaces for cold atoms \cite{gonzalez2015subwavelength} or solid state quantum emitters \cite{englund2010deterministic,gould2016efficient}.  Thus a variety of GaP nanophotonic devices have been fabricated and studied, including 1D \cite{schneider2018gallium} and 2D \cite{rivoire2009second,gan2012high,p1999strongly} photonic crystals, microdisks \cite{barclay2009chip,mitchell2014cavity,guilleme2017second,thomas2014waveguide,gould2016efficient,cambiasso2017bridging}, and strip waveguides \cite{thomas2014waveguide,gould2016efficient,schneider2018gallium}.

\textcolor{black}{Owing to its large nonlinearity and wide transparency, GaP is a promising material for integrated nonlinear photonics.} To fulfill this promise, crystalline GaP must be integrated onto a low-index material---ideally by a method compatible with waferscale production---then patterned into devices with sufficiently low propagation loss \textcolor{black}{to permit net optical gain}.  While observation of $\chi^{(2)}$ effects (e.g. second harmonic generation) in photonic crystal and microdisk resonators is an important step \cite{rivoire2009second,lake2016efficient}, these devices were realized by suspending the resonator in air, precluding the use of integrated waveguides and couplers.  Efforts to interface NV centers have produced techniques to bond GaP onto diamond ($n_0=2.4$) \cite{englund2010deterministic,gould2016efficient}; however, these are incompatible with large-scale wafer-level fabrication.  Moreover, in both cases, propagation loss was insufficient to observe strong $\chi^{(3)}$ effects.  Indeed, to date, the $\chi^{(3)}$ nonlinearity of GaP has only been measured with ultra-short light pulses at optical wavelengths \cite{liu2010three,martin2018nonlinear}, and promising applications based on the Kerr effect, such as supercontinuum  \textcolor{black}{\cite{skryabin2010colloquium,halir2012ultrabroadband}} and frequency comb \cite{del2007optical,kippenberg2011microresonator} (in particular soliton \cite{herr2014temporal}) generation, have yet to be explored.

\begin{figure*}[t!]
	\includegraphics[width=2\columnwidth]{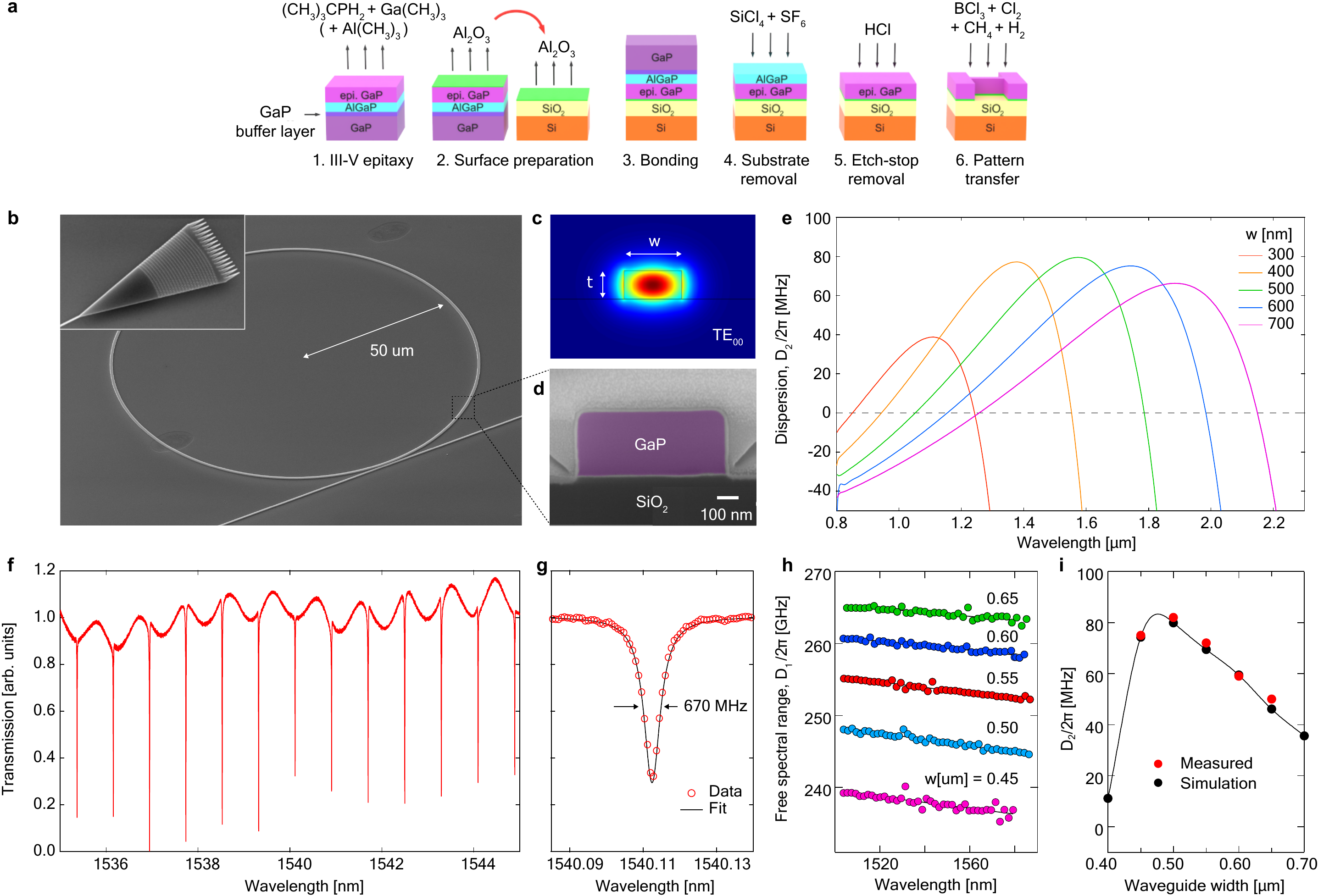}
	\caption{\color{black}\textbf{Gallium phosphide waveguide resonators}: (a) GaP-on-insulator fabrication process flow. (b) Scanning electron micrograph (SEM) of a 50-$\mu$m-radius ring resonator with integrated bus waveguide and grating couplers (inset).  Waveguide thickness and width are 300 nm and 500 nm, respectively. (c) Finite element simulation of the TE$_{00}$ waveguide mode. (d) SEM of the waveguide in cross-section. (e) Simulated resonator dispersion (\eqref{eq_2}) for various waveguide widths. (f) Measured transmission profile of \textcolor{black}{a silica-cladded racetrack resonator with free-spectral range $D_1$ = 100 GHz.  (g) Fit of the central peak to a Lorentzian with a linewidth of 0.67 GHz, corresponding a loaded $Q$ of $3\times 10^5$.} (h) Measurements of $D_1$ versus wavelength for 50-$\mu$m-radius rings with various waveguide widths.  (i) Comparison of measured (h) and simulated (e) dispersion.}
	\label{figure2}
	\vspace{-3mm}
\end{figure*}

Here we propose and implement a GaP-on-insulator platform for integrated nonlinear photonics. \textcolor{black}{Our approach \cite{schneider2018gallium}} makes use of direct wafer-bonding to integrate high-quality, epitaxially-grown GaP onto SiO$_2$. The large index contrast between GaP and SiO$_2$ ($n_0\approx 1.44$), in conjunction with highly anisotropic, low-roughness dry etching, allows us to realize single-mode strip waveguides with propagation losses as low as 1.2 dB/cm and nonlinearity parameters exceeding 200 W$^{-1}$m$^{-1}$ at 1550 nm. In this Letter, we engineer waveguide resonators to support Kerr frequency combs at telecommunications wavelengths.  We observe the first frequency comb in a GaP microresonator, with a threshold power as low as 3 mW, and with simultaneous formation of doubled combs at visible wavelengths.  \textcolor{black}{We moreover observe efficient Raman-scattered combs in a cladded resonator with reduced dispersion.}  

GaP-on-SiO$_2$ strip waveguides, as shown in \fref{figure2}, form the basis of our experiments.  To fabricate these devices, we \textcolor{black}{have developed} a novel III-V hybrid integration process based on direct wafer bonding \cite{schneider2018gallium}.  As  shown in \fref{figure2}a, the GaP device layer (300 nm thick) is first grown epitaxially, by metal-organic chemical vapor deposition (MOCVD), on a GaP substrate, with an intermediate Al$_{0.38}$Ga$_{0.62}$P layer included as an etch stop.  The growth wafer is then directly bonded to a Si wafer with a 2 $\mu$m thermal oxide, after depositing a thin (5 nm) layer of Al$_2$O$_3$ on both surfaces.  To remove the growth substrate, we use a combination of wet etching and a rapid, highly selective dry etch based on inductively-coupled plasma reactive ion etching (ICP-RIE) with SiCl$_4$ and SF$_6$ \cite{honl2018highly}.  The AlGaP layer is then removed with concentrated HCl, leaving the GaP device layer exposed for processing. Devices for this report were patterned by electron-beam lithography using HSQ as a negative resist. Patterns were then transferred by chlorine-based ICP-RIE \cite{suppinfo}. In a final step, 5 nm of Al$_2$O$_3$ is deposited onto the waveguide to mitigate photo-oxidation.

We employ single-crystal substrates and therefore epitaxial material with a non-centrosymmetric cubic (zincblende) crystal structure, resulting in a nonzero $\chi^{(2)}$ nonlinearity in addition to $\chi^{(3)}$.  Photons propagating in our waveguides therefore undergo both three-wave-mixing (TWM) and four-wave-mixing (FWM) under appropriate phase matching conditions. To observe these \textcolor{black}{nonlinear scattering processes}, it is necessary to pump the waveguide with sufficient optical power that the rate of \textcolor{black}{nonlinear scattering} exceeds that of linear loss (the condition for net parametric gain).  This condition is greatly facilitated by incorporating the waveguide into an optical cavity, such \mbox{that the circulating power is resonantly enhanced.}  

Waveguides were thus patterned into ring resonators. A typical 50-$\mu$m-radius ($R$) ring with integrated bus waveguide and grating couplers \cite{schneider2018gallium} is shown in \fref{figure2}a, based on a single-mode $300\times 500$ nm$^2$ waveguide.  Transmission measurements performed on cladded and uncladded resonators reveal that critically-coupled $Q$ factors as high as $2.5\times 10^5$ are achievable in both cases at C-band wavelengths.  The corresponding linear propagation loss, $\alpha \approx 1.2$ dB/cm, is attributed to sidewall roughness \cite{suppinfo}, and is on par with state-of-the-art AlGaAs \cite{pu2016efficient} and Si \cite{griffith2015silicon} waveguides of similar dimensions.  

\begin{figure}[t!]
	\includegraphics[width=1\columnwidth]{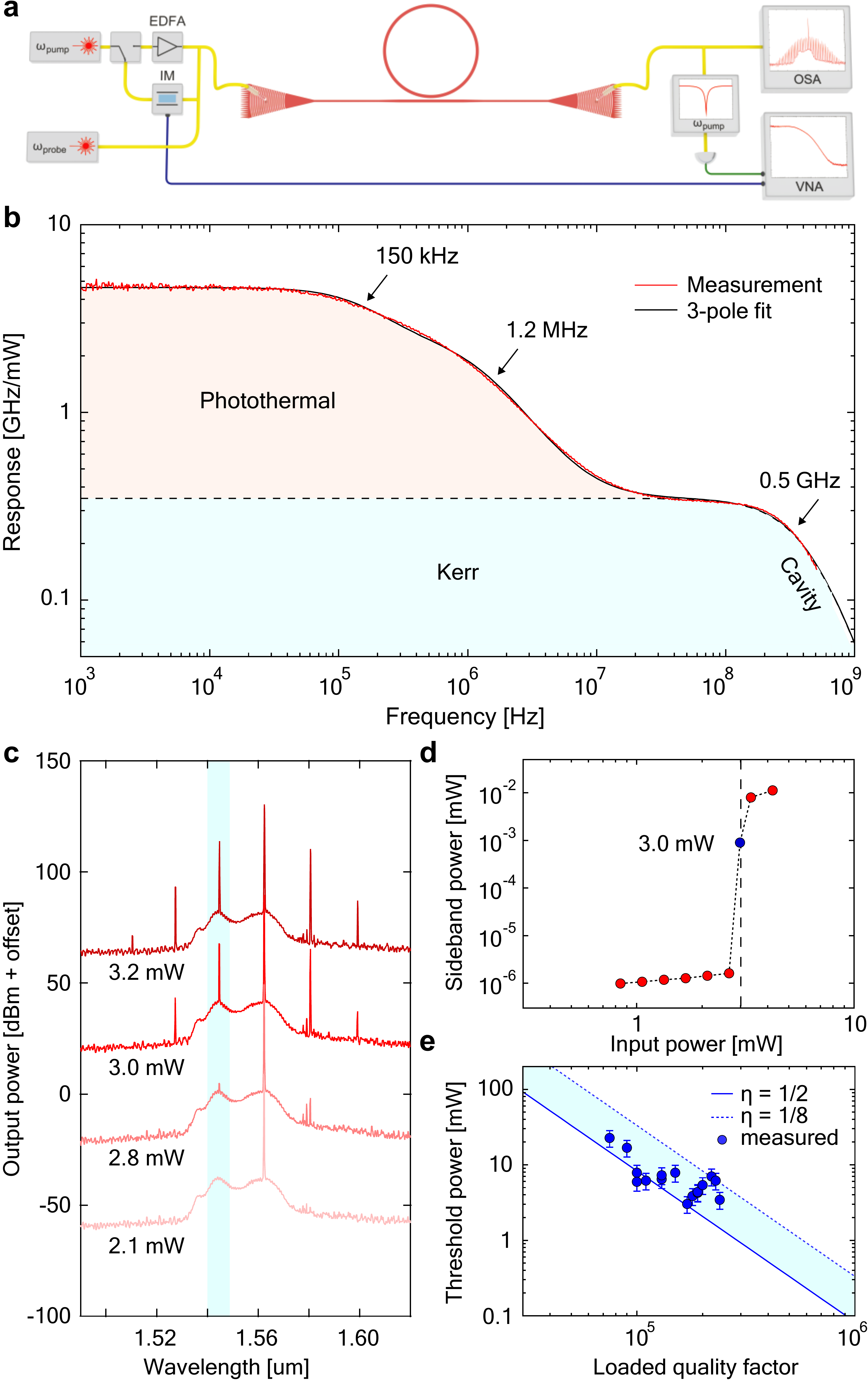}
	\caption{\color{black}\textbf{Linear and nonlinear response of a GaP micro-resonator.} (a) Setup for pump-probe response and frequency comb measurements \cite{suppinfo}. (b) Power-frequency response of a 50-$\mu$m-radius ring resonator. Plateaus at 0-100 kHz and 10-100 MHz are due to the photothermal and Kerr effect, respectively. Rolloff at 0.5 GHz is due to the finite cavity response time.  (c) Optical spectrum of the cavity output field as a function of input power.  Parametric oscillation (spontaneous FWM) is observed above a threshold value $P_\t{th}$. (d) Power in the left primary sideband versus input power, revealing $P_\t{th}\approx3\;\t{mW}$ for a $Q\approx2.5\cdot 10^5$ device. (e)  Measurements of $P_\t{th}$ versus $Q$ for various devices, compared to \eqref{eq:Pthres}.}
	\label{figure3}
	\vspace{-3mm}
\end{figure}

We next studied the $\chi^{(3)}$ nonlinearity of our waveguides, parameterized by the 
nonlinear propagation constant $\beta_\t{NL} = \gamma_\t{NL} P$ \cite{suppinfo}.  Here $\gamma_\t{NL}=\omega n_2 /A_\t{eff}c$ is the waveguide nonlinearity parameter, $n_2$ is the nonlinear refractive index (Kerr coefficient) of the waveguide core material, $A_\t{eff}$ is the effective core area \cite{suppinfo}, and $P$ is the guided power. To measure $\gamma_\t{NL}$, we used a modulation-transfer technique that allows separation of Kerr and photothermal nonlinearities based on their response times \cite{rokhsari2005observation}.  As illustrated in \fref{figure3}, an intensity-modulated pump field was injected into the cavity on resonance and a detuned probe field was used to record the frequency response of an adjacent cavity mode. At sufficiently high frequency, the response is dominated by the Kerr effect
\begin{equation} 
\delta\omega_\t{probe}(\Omega)\approx  \frac{8\eta Qc}{V_\t{eff}}\frac{n_2}{n_\t{g}^2}\delta P_\t{pump}(\Omega)
\end{equation}
where $V_\t{eff}=2\pi R A_\t{eff}$ is the effective resonator mode volume, $n_\t{g}$ is the waveguide group index, $\Omega$ is the modulation frequency, $P_\t{pump}$ is the injected pump power, and $\eta\in[0,1]$ is the cavity impedance matching factor  \cite{suppinfo}. \textcolor{black}{Normalizing} the response curve to the DC photothermal frequency shift, we infer  \textcolor{black}{$\gamma_\t{NL}=2.4(6)\times10^2\,\t{W}^{-1}\t{m}^{-1}$ and $n_2=1.1(3)\times 10^{-17}\,\t{m}^{2}/\t{W}$}. (For details see \cite{suppinfo}.) Somewhat lower values for $n_2$ have been obtained using pulsed measurements \cite{liu2010three,martin2018nonlinear}.  
A  comparison to other integrated photonics materials is made in \fref{figure1}.  

At powers above the parametric oscillation threshold (note that the formula below accounts for dispersion \cite{kippenberg2004kerr,chembo2010modal,suppinfo}) 
\begin{equation}\label{eq:Pthres}
P_\t{th} \approx  \frac{\pi}{4\eta}\frac{V_\t{eff}}{\lambda Q^2}\frac{n_g^2}{n_2},
\end{equation}
recirculation of scattered photons can give rise to cascaded FWM into multiple cavity modes, forming a frequency comb.  Such microresonator frequency combs \cite{kippenberg2011microresonator} have become a front-line research topic in recent years due to their broad applicability in telecommunications and metrology \cite{kippenberg2011microresonator}, particularly with the advance of low-noise, soliton frequency combs.  Among the growing number of material platforms used to realize microresonator frequency combs (\fref{figure1}), the only highly nonlinear platforms are AlGaAs- \cite{pu2016efficient} and Si-on-insulator \cite{griffith2015silicon}, both of which suffer from TPA at telecommunication wavelengths.  In addition to its wide transparency window, generating microresonator combs in GaP holds intrigue due to its $\chi^{(2)}$ nonlinearity and piezo-electricity, which might permit, for example, self-referencing by in-situ frequency doubling \cite{xue2017second} or electronic control \cite{griffith2015silicon} of comb dynamics.  

\begin{figure*}[t!]
	\includegraphics[width=2.05\columnwidth]{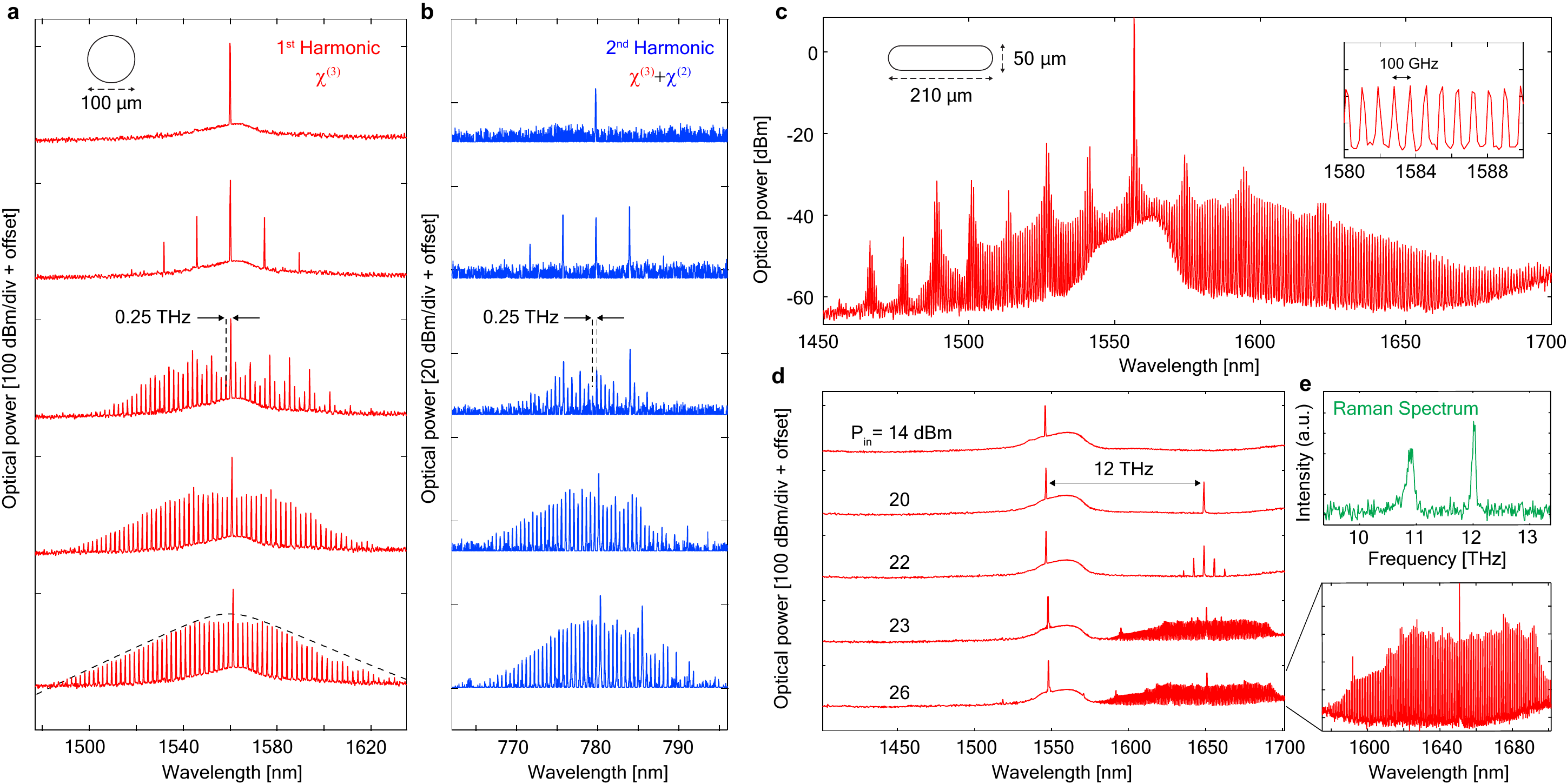}
	\caption{\color{black}\textbf{Gallium phosphide microresonator frequency combs.} (a) Frequency comb generation in a 50-$\mu$m-radius ring resonator.  From top to bottom, the laser power is fixed and the laser-cavity detuning is incrementally reduced.  A sech-squared envelope (dashed black) is overlaid on the last spectrum.  (b)  Simultaneous measurements of frequency-doubled combs on the same device. (c) Broadband frequency comb generated in a cladded racetrack resonator with a FSR of 100 GHz. (d) Raman-shifted frequency comb observed on a different mode of the same device. \mbox{(e) Raman spectrum of the GaP growth substrate.}}
	\label{figure4}
	\vspace{-3mm}
\end{figure*}

In \fref{figure4} we present the first microresonator frequency combs in GaP.  To realize this advance, a crucial consideration is waveguide dispersion, which regulates the phase-matching of FWM-scattered photons:  Specifically, to compensate self- and cross-phase modulation \cite{herr2013solitons}, efficient FWM requires that the waveguide exhibit anomalous group velocity dispersion, $\t{GVD}\equiv(1/c)\partial n_\t{g}/\partial\omega<0$,
which manifests as dispersion of the resonator free spectral range, $D_1=c/(n_g R)$:
\begin{equation}\label{eq_2}
D_2(\omega)\equiv D_1\frac{\partial D_1}{\partial \omega} \approx -D_1^2\frac{c}{n_\t{g}}\t{GVD}>0.
\end{equation}
Like most materials, in bulk form, GaP exhibits normal GVD in its transparency window. To overcome this material dispersion, it is necessary to tune the waveguide dimensions to realize strongly anomalous geometric GVD. 

Waveguides were therefore dispersion-engineered by fine-tuning their width.  Using the Sellmeier equation to estimate the material GVD of GaP, finite element simulations predict that a broadband anomalous GVD window centered at 1550 nm can be achieved for a single-mode waveguide with a thickness of 300 nm and a width of 500 nm (\fref{figure2}e).  To confirm this model, ring resonators were patterned with waveguide width varied from 450 to 650 nm. Measurements of FSR were then carried out (\fref{figure2}h) as a function of wavelength. Results were found to be in good agreement with \eqref{eq_2} (\fref{figure2}i).

The onset of parametric oscillation in a 50-$\mu$m radius, $550$-nm-wide ring resonator is shown in \fref{figure3}c,d.  For this measurement the pump laser was passed through an erbium-doped fiber amplifier and the resonator output was directed to an optical spectrum analyzer \cite{suppinfo}.  The laser was stabilized at a small blue-detuning using the photothermal self-lock.  While increasing the power (and simultaneously compensating the photothermal frequency shift), sideband generation was observed at a threshold of $P_\t{th}\approx 3$ mW, for a near-critically coupled mode at 1560 nm with $Q\approx 2\times 10^5$.  Similar measurements were carried out for modes with $Q=(0.7-2.5)\times10^5$ by varying the operating wavelength and bus-waveguide separation (\fref{figure3}e).  Comparing this set of measurements to \eqref{eq:Pthres}, we infer \textcolor{black}{$n_2\approx 1.2(5)\times10^{-17}\,\t{m}^2/\t{W}$, in good agreement with the value inferred from the linear response measurement.}

\textcolor{black}{Using input powers greater than $P_\t{th}$, broadband frequency comb generation was observed in numerous devices.  An example is shown in \fref{figure4}a,} in which the input power was fixed at 1 W and the laser-cavity detuning was incrementally reduced.  Typical behavior is shown: Initially sub-combs separated by multiple FSR are formed (consistent with phase-matching for the far-detuned pump \cite{herr2012universal}).  At smaller detunings, interleaving of sub-combs results in a ``full" comb with a characteristic sech-squared envelope, here with a 3 dB width of 1.4 THz, limited by the large magnitude of the resonator dispersion \cite{herr2014temporal}. We note that although the sech-squared envelope is consistent with coherent comb (soliton) formation \cite{herr2014temporal}, excess radiofrequency noise in the optical power spectrum \cite{suppinfo} implies that the combs shown are not fully coherent \cite{herr2012universal}.  A possible reason for the resistance to soliton formation is perturbation of the cavity dispersion due to mode splitting \cite{herr2014mode}.

In tandem with FWM, the $\chi^{(2)}$ nonlinearity of GaP enables frequency combs generated in the C-band to be simultaneously doubled to visible wavelengths. 
Frequency-doubled combs produced simultaneously with the primary combs in \fref{figure4}a are shown in \fref{figure4}b.  As visible light is not efficiently coupled from the ring to the bus waveguide \cite{guo2016second}, for these measurements the output fiber was positioned above the ring.  An auxiliary high-resolution visible OSA  was used \cite{suppinfo}, revealing as many as 50 comb lines (limited by the OSA sensitivity) for the smallest laser-cavity detuning .  

Finally, we examined the role of Raman scattering in our devices.  It is well known that stimulated Raman scattering (SRS) competes with FWM in waveguides with normal or weak dispersion \cite{min2005controlled}.  We observe this effect in waveguide resonators cladded in silica, which reduces $D_2$ by an order of magnitude and displaces the center of the anomalous window to approximately 1650 nm. 
FWM and SRS in a 100 GHz cladded racetrack resonator with $Q\approx2\times10^5$ is shown in \fref{figure4}c,d.  For a subset of modes, broadband ($>200$ nm) frequency combs centered at the pump wavelength are generated by FWM.  For others, FWM is preceded by efficient Raman lasing (20 dBm threshold, $10\%$ conversion efficiency) with a Stokes frequency of 12 THz.  The selectivity between FWM and SRS, and the absence of SRS in resonators with smaller radii, is likely due to the narrowness of the Raman transition, inherent to the crystalline material.  Independent Raman spectroscopy of the GaP growth substrate (\fref{figure4}e) confirms that the 12 THz transition is less than 100 GHz wide.  

In summary, we have explored GaP-on-insulator as a platform for nonlinear photonics, using microresonator frequency comb generation was as an illustrative example. Operating at telecommunications wavelengths, propagation losses as low as 1.2 dB/cm were observed for single-mode strip waveguides with effective areas of $0.2\,\mu\t{m}^2$.  Exploiting precision control over sidewall dimensions, waveguides were dispersion-engineered to support efficient FWM.  Resonators formed from these waveguides were found to exhibit parametric oscillation with as little as 3 mW injected power, followed by formation of broadband ($>100$ nm) frequency combs with comb spacing ranging from 100 to 250 GHz, depending on resonator geometry.  In conjunction with pump-probe measurements, a direct measurement of the Kerr coefficient of GaP was made at telecommunication wavelengths: $n_2=1.2(5)\cdot10^{-17}\,\t{m}^2/\t{W}$.  Also observed were frequency-doubled combs and Raman-shifted frequency combs. Looking forward, numerous applications of GaP-on-insulator nonlinear photonics are expected to exhibit high performance.  Reduction of waveguide sidewall roughness and enhanced dispersion engineering might allow, for example, soliton comb formation, mid-IR frequency combs, and ultra-broadband supercontinuum generation. We conclude with Table 1, in which the linear and nonlinear properties of GaP microresonators are compared to that of contemporary platforms for Kerr frequency comb generation.

\begin{table}\label{table1}
	\begin{tabular}{c|c|c|c|c|c|c|c|c}
		material & $n_0$ & $n_2$ & $\chi^{(2)}$ & $\lambda_\t{TPA}$ & $A_\t{phys}$ & $D_1$ & $Q$ & $P_\t{th}$ \\ 
		\hline &
		& $10^{-18}\,\frac{\t{m}^2}{\t{W}}$ & $\frac{\t{pm}}{\t{V}}$ & $\t{nm}$ & $\mu\t{m}^2$ & $\t{GHz}$ & $10^6$ & $\t{mW}$\\ 
		\hline 
		Si & 3.5 & 4 & -- & 2250 & -- & -- & -- & -- \\
		\hline 
		Al$_{.83}$Ga$_{.17}$As & 3.3 & $26$ & 120 & 1540 &  0.20 & 995 & 0.1 & 3\\ 
		\hline
		\textcolor{red}{GaP} & \textcolor{red}{3.1} & \textcolor{red}{$12(5)$} & \textcolor{red}{82} & \textcolor{red}{1100} & \textcolor{red}{0.15} & \textcolor{red}{250} & \textcolor{red}{0.2} & \textcolor{red}{3}\\ 
		\hline 
		Diamond & 2.4 & 0.082 & -- & 450 & 0.81 & 925 & 1 & 20 \\
		\hline
		AlN & 2.1 & 0.23 & 0.43 & 440 & 2.3 & 435 & 0.8 & 200 \\  
		\hline
		Si$_3$N$_4$ & 2.0 & 0.25 & -- & 460 & 1.8 & 200 & 36 & 0.3 \\
		\hline
		Hydex & 1.7 & 0.12 & -- & 280 & 2.2 & 200 & 1 & 50 \\
		\hline
		SiO$_2$ & 1.4 & 0.022 & -- & 280 & $\sim\,$30 & 33 & 270 & 1 \\
		\hline
	\end{tabular}
	\caption{Properties of current integrated microresonator frequency comb platforms operating at $\lambda\sim1.55\,\mu\t{m}$. $\lambda_\t{TPA}$ and $A_\t{phys}$ correspond to the cutoff wavelength for two-photon absorption and the physical waveguide area, respectively. \textcolor{black}{Values were taken from \cite{dinu2003third} (Si), \cite{pu2016efficient} (AlGaAs), \cite{hausmann2014diamond} (Diamond), \cite{jung2013optical} (AlN), \cite{ji2017ultra} (Si$_3$N$_4$), \cite{razzari2010cmos} (Hydex), and \cite{pu2016efficient} (SiO$_2$).} }
\end{table}

\section*{Acknowledgements}
We gratefully acknowledge Pol Welter, Yannick Baumgartner, Herwig Hahn, Lukas Czornomaz, Ute Drechsler, Daniele Caimi, and Antonis Olziersky for their valuable contributions to development of the GaP-on-insulator platform.  We also thank Maxim Karpov and Tobias Herr for useful discussions about frequency comb generation. \textcolor{black}{This work was supported by the European Union's Horizon 2020 Program for Research and Innovation under grant agreement No. 722923 (Marie Curie H2020-ETN OMT) and No. 732894 (FET Proactive HOT).} All samples were fabricated at the Binnig and Rohrer Nanotechnology Center (BRNC) at IBM Research, GmbH.

\section*{Contributions}
K.S. and P.S. developed the GaP-on-insulator platform with support from S.H..  S.H. fabricated all devices used in the reported experiments, took SEM and AFM images, and performed Raman measurements. D.J.W. conducted microresonator experiments and analyzed data with support from K.S., S.H., and P.S..  M.A. provided numerical models and general theory support. D.J.W. wrote the manuscript with support from all co-authors.  T.J.K. guided the investigation of frequency combs. P.S. conceived and oversaw the project.


%

\clearpage

\onecolumngrid
\setcounter{equation}{0}
\setcounter{figure}{0}
\setcounter{section}{0}

\renewcommand{\thefigure}{S\arabic{figure}}
\renewcommand{\theequation}{S\arabic{equation}}

\renewcommand{\abstractname}{}

\vspace{1cm}
\begin{center}
	\large\textbf{Supplementary Information for ``Gallium Phosphide Nonlinear Photonics"}
\end{center}

\begin{center}
\normalsize Dalziel J. Wilson$^{1,2}$, Katharina Schneider$^1$, Simon H\"{o}nl$^1$, Miles Anderson$^2$, Tobias J. Kippenberg$^2$, and Paul Seidler$^1$
\end{center}

\begin{center}
	$^1$\emph{IBM Research --- Zurich, Sa\"{u}merstrasse 4, 8803 R\"{u}schlikon, Switzerland}\\
    $^2$\emph{Institute of Physics (IPHYS), {\'E}cole Polytechnique F{\'e}d{\'e}rale de Lausanne, 1015
		Lausanne, Switzerland}\\
	(Dated: August 10,2018)
\end{center}

\begin{quote}
\hspace{0.5cm}Information supplementary to the main text is here provided, 
including definitions of the nonlinear refractive index, derivation of the Kerr frequency shift, derivation of the power threshold for frequency comb generation,  and details about the pump-probe response measurement.  Particular attention is paid to chromatic dispersion, which requires use of the group index in place of refractive index in key instances.  Fabrication is also elaborated upon.
\end{quote}
\vspace{0.3cm}

\twocolumngrid

\section{Nonlinear propagation constant}
The propagation constant $\beta(\omega)$ of a waveguide (or equivalently, its effective refractive index $n(\omega)$) is given by
\begin{equation}
s_\t{out}(\omega)=s_\t{in}(\omega)e^{i\beta(\omega)x}\equiv s_\t{in}(\omega)e^{i\omega n(\omega) x/c}
\end{equation}
where $s_\t{in(out)}$ and $\omega$ are the amplitude and frequency of the waveguide input (output) field, respectively, $c$ is the speed of light in vacuum, and $x$ is the propagation distance.

If the waveguide core material is $\chi^{(3)}$-nonlinear, then the effective index can be expressed in the form
\begin{equation}\label{eq_nonlinearindex}
n(\omega)=n_\t{0}(\omega)+n_2(\omega)\frac{P}{A_\t{eff}(\omega)}
\end{equation}
where $n_2$ is the nonlinear refractive index of the core material, $n_0$ is the ``cold" effective index, $A_\t{eff}$ is the effective area of the waveguide mode \cite{agrawal2000nonlinear_B}, and $P$ is the guided power.

As described in the main text, we estimate $n_2$ by combining a measurement of the nonlinearity parameter
\begin{equation}\label{eq_gammaNL}
\gamma_\t{NL}\equiv\frac{\partial\beta}{\partial P} = \frac{\omega n_2}{A_\t{eff} c}.
\end{equation}
with a numerical simulation of $A_\t{eff}$.  The latter is obtained using approximate expression \cite{agrawal2000nonlinear_B}
\begin{equation}\label{eq_Aeff}
A_\t{eff}\approx\frac{(\int I dA)^2}{\int I^2dA},
\end{equation}
where $I$ is the transverse intensity profile of the guided field and $\int dA$ is an area integral extending outside the core and over all space.  Simulations of $n_0$ and $A_\t{eff}$ for the TE$_{00}$ mode of a 300-nm-thick GaP strip waveguide are shown in Fig. S1, as a function of waveguide width and operating wavelength.  

\section{Waveguide resonator}

A resonator may be formed by closing the waveguide onto itself (forming a ring, in our case). Its resonance frequencies $\omega_m$ are given by
\begin{equation}\label{eq_resonancecondition}
\beta(\omega_m)L=2\pi m
\end{equation}
where $m$ is the mode index and $L$ is the physical length of the resonator. ($L = 2\pi R$ for a ring with physical radius $R$.)  
\subsection{Free spectral range and group index}
Accounting for chromatic dispersion ($d n/d\omega\ne 0$), the resonator free spectral range
\begin{equation}
D_{1}^{(m)} \equiv \frac{d\omega_m}{d m}
\end{equation} 
is given from \eqref{eq_resonancecondition} by
\begin{subequations}\label{eq_FSRdefinition}
	\begin{align}
	\frac{d\beta(\omega_m)}{dm} &= \frac{\partial\beta(\omega_m)}{\partial\omega_m}\frac{d\omega_m}{dm} = \frac{2\pi }{L}\\
	\rightarrow D_{1}^{(m)}&=\frac{2\pi c}{n_\t{g}(\omega_m)L}
	\end{align}
\end{subequations}
where 
\begin{equation}\label{eq_ngdefinition}
n_\t{g}(\omega)\equiv c\frac{d\beta}{d\omega}= n(\omega)+\omega\frac{\partial n}{\partial\omega}
\end{equation}
is the waveguide group index.

Eq. \ref{eq_FSRdefinition}b was used to derive the group index from resonator transmission measurements described in the main text.  Because of large material and geometric dispersion, the difference between $n_\t{g}$ and $n_0$ is significant for our GaP waveguides, e.g. $\{n_\t{g},n_0,n_0^\t{(mat)}\}=\{3.8,2.2,3.1\}$ (evaluated at 1550 nm) for a $300\times500\,\t{nm}^2$ waveguide, where $n_0$ is determined from finite element simulation and $n_0^\t{(mat)}$ is the bulk (material) index obtained, e.g., from the Sellmeier equation \cite{wei2018temperature_B} .
%
\subsection{Kerr frequency shift}

Resonance frequencies of a nonlinear waveguide resonator in general depend on the circulating power $P$.  Using Eqs. \ref{eq_gammaNL} and \ref{eq_FSRdefinition} and neglecting dispersion of $n_2$ and $A_\t{eff}$, one finds
\begin{subequations}\label{eq_dfdP}
	\begin{align}
	\frac{d\beta(P,\omega_m)}{dP}&=\frac{\partial\beta}{\partial P}+\frac{\partial\beta(\omega_m)}{\partial\omega_m}\frac{d \omega_m}{dP}=0\\
	\rightarrow \frac{d\omega_m}{d P}&=-\frac{\omega_m n_2}{n_\t{g} A_\t{eff}}.
	\end{align}
\end{subequations}

Eq. \ref{eq_dfdP} was used to derive Eq. 1 in the main text, using cavity input-output relations to relate $P$ to the input power $P_\t{in}$ (\eqref{eqn_steadystateHOTcrit}), and multiplying the right hand side by 2 to account for cross phase modulation \cite{agrawal2000nonlinear_B}, viz.

\begin{equation}\label{eq_dfdP_xpm}
\frac{d\omega_{m'\ne m}}{d P}=2	\frac{d\omega_{m}}{d P}
\end{equation} 

\begin{figure}[t!]
	\includegraphics[width=1\columnwidth]{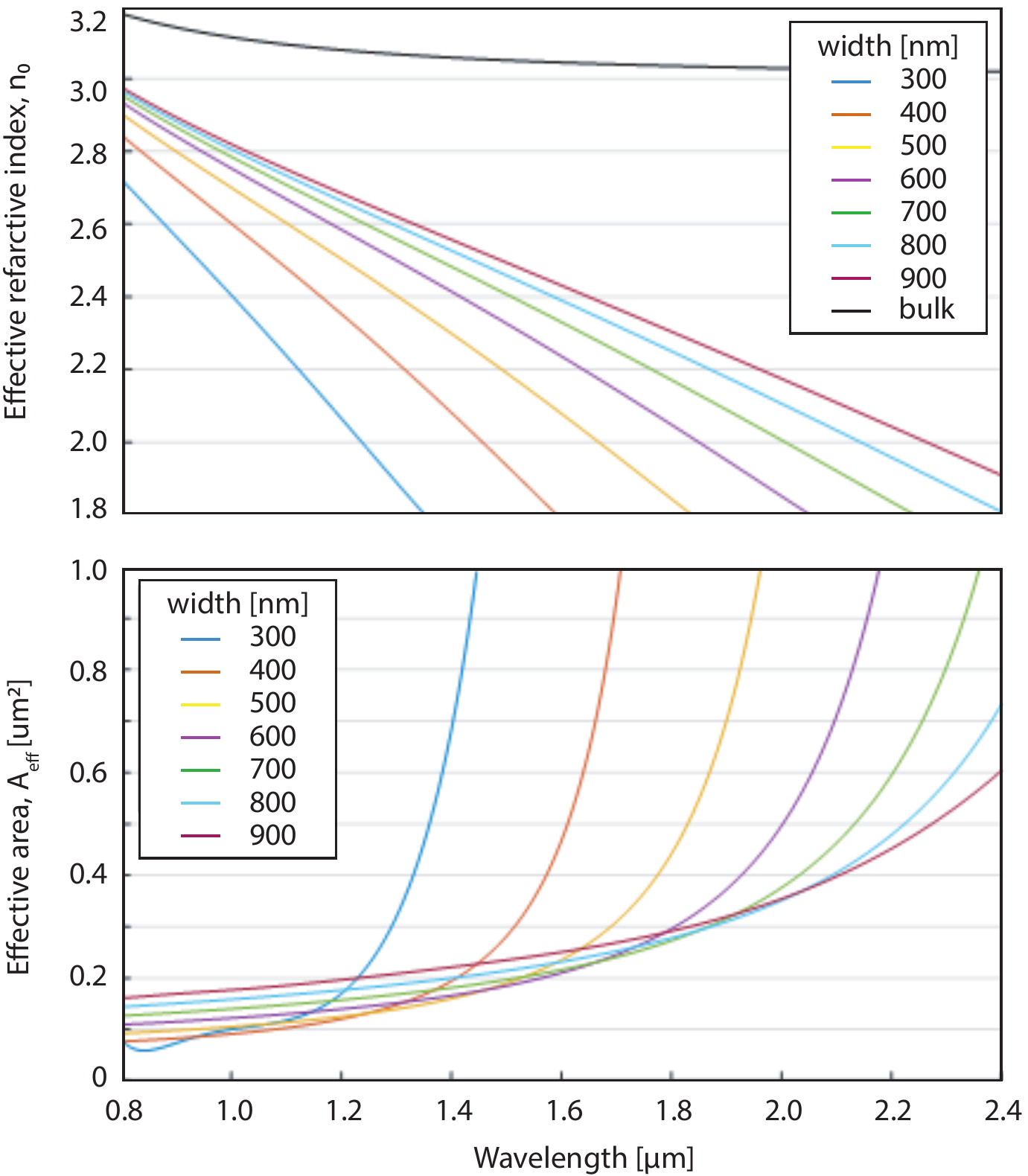}
	\caption{\textbf{Effective index and area of uncladded GaP-on-SiO$_2$ strip waveguides.} (a) Finite element simulation of $n_0(\lambda)$ for 300-nm-thick waveguides of various widths, excited in the TE$_{00}$ mode.  Also shown is the bulk index, $n_0^{(\t{mat})}(\lambda)$, obtained from the Sellmeier equation \cite{wei2018temperature_B}. (b) Simulation of $A_\t{eff}$ for the same set of waveguides, using \eqref{eq_Aeff}.}
	\label{fig_1}
\end{figure}

\subsection{Open system dynamics}

The Kerr frequency shift (\eqref{eq_dfdP}) gives rise to rich dynamics when opening the resonator to an external drive. Adopting standard input-output formalism, one arrives at the following nonlinear equation of motion for the circulating field \cite{agrawal2000nonlinear_B}
\begin{subequations}\label{eqn_cavityeqofmotion}\begin{align}
	\dot{a} &= -(\kappa/2-i(\Delta_0+G|a|^2))a+\sqrt{\kappa_\t{ex}}s_\t{in}\\
	s_\t{out}&=s_\t{in}-\sqrt{\kappa_\t{ex}}a.
	\end{align}
\end{subequations}
Here $a$ and $s_\t{in(out)}$ are the amplitude of the circulating and input(output) fields, normalized so that $|a|^2= P\tau_\t{rt}$ is the circulating energy and $|\bar{s}_\t{in(out)}|^2 = P_\t{in(out)}$ is the input(output) power, $\tau_\t{rt}\equiv 2\pi/D_1$ is the cavity round-trip time, $\Delta_0=\omega_\t{m}-\omega_0$ is the detuning of the input field from the \emph{cold} resonance $(|a|^2=0)$, $\kappa=\kappa_\t{ex}+\kappa_0$ is the total resonator damping rate, including contributions from  internal loss, $\kappa_0$, and coupling to the bus waveguide $\kappa_\t{ex}$, and $G|a|^2$ is the nonlinear frequency shift.  From \eqref{eq_dfdP}, coupling term $G$ is given by
\begin{equation}\label{eq_G}
G = \frac{d\omega_m}{dP}\frac{1}{\tau_\t{rt}}=\frac{\omega_m n_2 c}{n_\t{g}^2 V_\t{eff}}
\end{equation}
where $V_\t{eff}=A_\t{eff}L$ is the effective mode volume.

Defining $\Delta \equiv \Delta_0+G|a|^2$ as the detuning of the laser from the \emph{hot} resonance, steady state solutions to \eqref{eqn_cavityeqofmotion} may be expressed in the familiar form
\begin{subequations}\label{eqn_steadystateHOT}
	\begin{align}
	P&=\frac{2\eta \mathcal{F}}{\pi}\mathcal{L}(\Delta)P_\t{in}\\
	P_\t{out}&=(1-4\eta(1-\eta)\mathcal{L}(\Delta))P_\t{in}
	\end{align}
\end{subequations}
where $\eta = \kappa_\t{ex}/\kappa$ is the bus-resonator impedance matching factor, $\mathcal{F}=D_1/\kappa$ is the resonator finesse and
\begin{equation}
\mathcal{L}(\Delta)=\frac{1}{1+4\Delta^2/\kappa^2}
\end{equation}
is the normalized resonator transmission.

For example, in the case of a resonant input field ($\Delta=0$), \eqref{eqn_steadystateHOT} simplifies to
\begin{subequations}\label{eqn_steadystateHOTcrit}
	\begin{align}
	P^{(0)}&=\frac{2\eta\mathcal{F}}{\pi}P_\t{in}\\
	P_\t{out}^{(0)}&=(1-2\eta)^2P_\t{in},
	\end{align}
\end{subequations}
with maximum (minimum) values for the circulating (output) power occurring for critical coupling, $\eta = 1/2$. 


%
%

\subsection{Modulation instability}

\begin{figure*}[t!]
	\includegraphics[width=1.33\columnwidth]{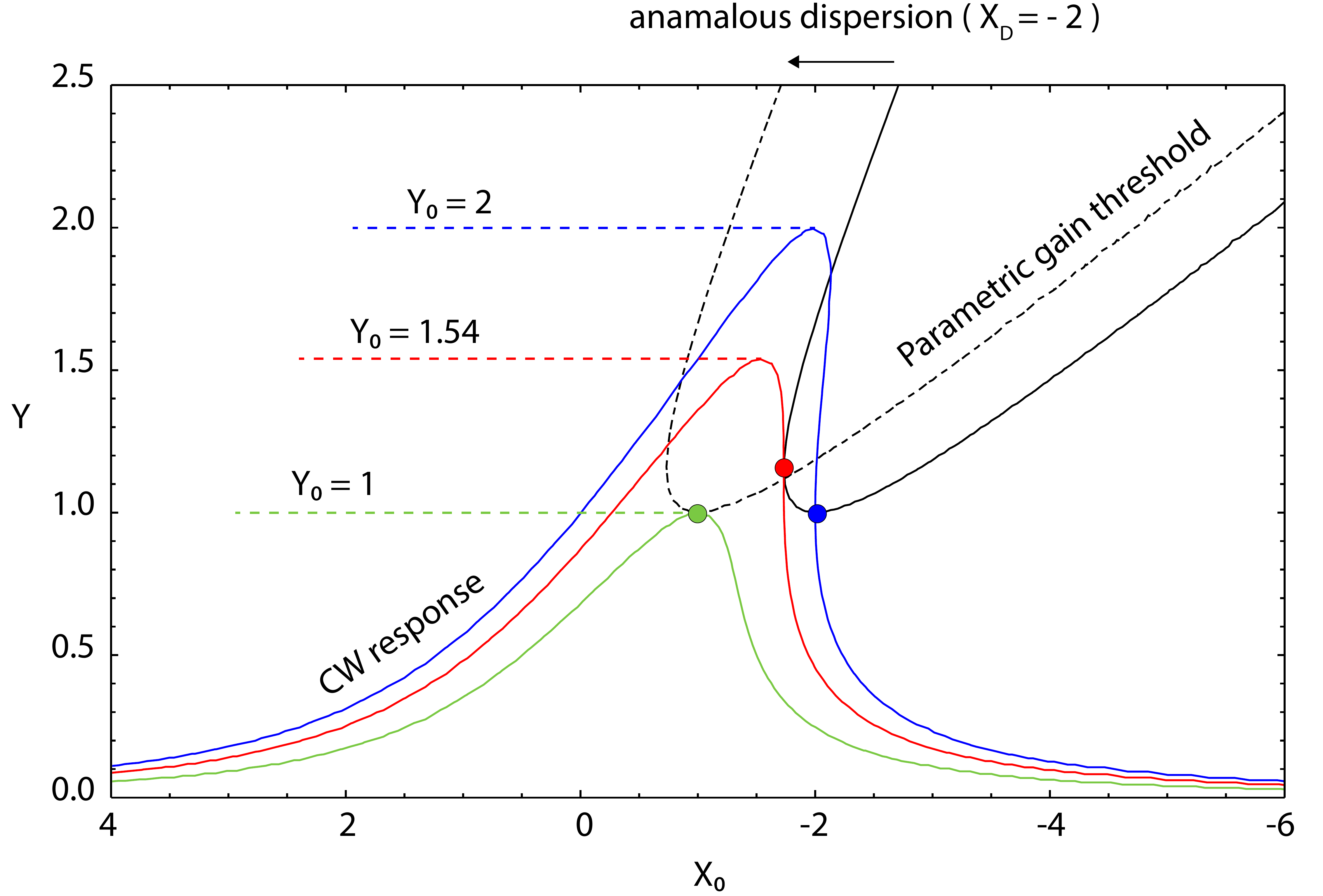}
	\caption{\textbf{Visualization of the threshold condition for parametric oscillation.}  Solid blue, red, and green lines: Steady-state response of the circulating power ($Y$) to a change in the frequency of the drive field ($X_0$), following \eqref{eq_duffing}. Normalized units $\{X_0,Y_0,Y\}$ described in \eqref{eq_XY} are used.  Response curves for input powers $Y_0 =\{1,1.54,2\}$ are shown. Solid and dashed black lines: Threshold condition for parametric oscillation in the presence of dispersion ($X_D$), following \eqref{eq_oscillationthreshold}.  Solutions for zero dispersion ($X_D=0$, solid) and anomalous dispersion ($X_D = -2$, dashed) are shown.  Blue, red and green points highlight three commonly cited oscillation thresholds: Minimal circulating power in the absence of dispersion (blue point), minimal input power in the absence of dispersion (red point, corresponding to the MI threshold, \eqref{eq_PthMI}), and minimal input (and circulating) power in the presence of anomalous dispersion (green point, corresponding to Eq. 3 in the main text).}
	\label{fig_2}
\end{figure*}

For sufficiently large circulating powers $P$, taking the partial derivative of \eqref{eqn_steadystateHOT} with respect to $\Delta$ does not correctly predict the response $\delta P$ to an external frequency modulation $\delta\Delta_0$.  For this, one must take into account the power dependence of the detuning $\Delta(P)$. Indeed, for sufficiently large $P$ the response to an external modulation can diverge.  This ``modulation instability'' (MI) may be understood as the physical origin of frequency comb generation \cite{matsko2005optical_B,chembo2010modal_B}. 

Here we derive the threshold condition for modulation instability by graphically solving the explicitly power-dependent form of \eqref{eqn_steadystateHOT}
\begin{equation}\label{eq_duffingraw}
P\left(1+4(\Delta_0+G\tau_\t{rt} P)^2/\kappa^2\right)=P^{(0)}
\end{equation}

Eq. \ref{eq_duffingraw} describes the familiar steady-state response of a driven duffing oscillator. It is convenient to rewrite it as
\begin{equation}\label{eq_duffing}
Y(1+(X_0+Y)^2)=Y_0
\end{equation}
by introducing the normalized coordinates
\begin{subequations}\label{eq_XY}
	\begin{align}
	X_0 &= \frac{2\Delta_0}{\kappa}\\
	Y &= \frac{2(\Delta-\Delta_0)}{\kappa}=\frac{2G\tau_\t{rt}P}{\kappa}\\
	Y_0 &=\frac{2G\tau_\t{rt}P^{(0)}}{\kappa}=\frac{8\eta G}{\kappa^2}P_\t{in}.
	\end{align}
\end{subequations}
$X_0$, $Y$, and $Y_0$ are proxies for the input field detuning, circulating power, and input power, respectively. Specifically, they correspond to the cold detuning ($X$), Kerr frequency shift ($Y$), and maximum (resonant) Kerr frequency shift ($Y_0$), each scaled to the cavity resonance half-width, $\kappa/2$. 

Plotting solutions to \eqref{eq_duffing} for different input powers $Y_0$  (Fig. \ref{fig_2}), one observes that beyond a certain threshold, $Y_0^\t{MI}\approx 1.54$, there always exists at least one cold detuning $X_0$ for which the response of the circulating power to a input frequency modulation diverges, $\partial Y/\partial X_0=\infty$.  Using \eqref{eq_XY}c and \eqref{eq_G}, one finds that this input power, the power threshold for MI, is given by
\begin{equation}\label{eq_PthMI}
P_\t{in}^\t{MI} = Y_0^\t{MI}\frac{\kappa^2}{8\eta G} \approx 1.54\frac{\pi}{4\eta}\frac{n_\t{g}^2}{n_2}\frac{V_\t{eff}}{Q^2\lambda}.
\end{equation} 

One can likewise identify a minimal cold detuning for MI:
\begin{equation}\label{eq_detuningMI}
\Delta_0^\t{MI}=\frac{\kappa}{2}X^\t{MI}=\frac{\sqrt{3}}{2}\kappa
\end{equation}

\subsection{Threshold for frequency comb generation}

MI can be connected to frequency sideband generation --- and thereby frequency comb generation --- by seeking solutions to \eqref{eqn_cavityeqofmotion} of the form \cite{matsko2005optical_B,chembo2010modal_B}
\begin{equation}
a(t)=a_0 +\left(a_1e^{\lambda t}e^{-i\Omega t}+c.c.\right)
\end{equation}
where $\{\Omega,\lambda\}\in\Re$ and $\omega_m\gg\Omega \gg \kappa$. Here $\Omega$ corresponds the offset frequency of the first sideband generated by parametric oscillation and $\lambda$ corresponds to the net gain of the frequency conversion process.  The threshold condition for sideband generation (parametric oscillation) is given by $\lambda = 0$.  A careful analysis reveals that this condition is satisfied if \cite{chembo2010modal_B,herr2014temporal_B,herr2013solitons_B}
\begin{equation}\label{eq_oscillationthreshold}
Y^2-(X_0+X_D+2Y)^2=1
\end{equation}
where $X_D = 2D_1/\kappa\cdot\partial D_1/\partial\omega\equiv2D_2/\kappa$ is the normalized resonator dispersion.

Solutions to the parametric oscillation threshold condition (\eqref{eq_oscillationthreshold}) and steady-state nonlinear response (\eqref{eq_duffing}) are plotted for different dispersions and input powers, respectively, in \fref{fig_2}.  Intersections between these curves give the input power threshold ($Y_0^\t{th}$) for frequency comb generation. In absolute units:
\begin{equation}\label{eq_Pth}
P_\t{in}^\t{th} = Y_0^\t{th}\frac{\pi}{4\eta}\frac{n_\t{g}^2}{n_2}\frac{V_\t{eff}}{Q^2\lambda}\ge\frac{\pi}{4\eta}\frac{n_\t{g}^2}{n_2}\frac{V_\t{eff}}{Q^2\lambda}.
\end{equation}

Three commonly adopted thresholds are highlighted in \fref{fig_2} by solid circles: minimal circulating power in the absence of dispersion (blue circle, $\{X_D^\t{th},X_0^\t{th},Y_0^\t{th}\}=\{0,-2,1\}$), minimal input power in the absence of dispersion (red circle, $\{X_D^\t{th},X_0^\t{th},Y_0^\t{th}\}=\{0,-\sqrt{3},1.54\}$), and minimal input \emph{and} circulating power in the presence of anomalous dispersion (green circle $\{X_D^\t{th},X_0^\t{th},Y_0^\t{th}\}=\{-2,-1,1\}$).  The second threshold is equivalent to the MI threshold (\eqref{eq_PthMI}-\eqref{eq_detuningMI}).  The third gives the absolute minimum input power for frequency comb generation (\eqref{eq_Pth}), occurring for an anomalous dispersion of $D_2 = -\kappa$ and a cold-detuning of $\Delta_0 = \kappa/2$.
The latter, stricter threshold is cited in the main text (Eq. 3) and used to infer $n_2$ from measurements shown in Fig. 3.  

We remark that while \eqref{eq_Pth} has been widely used to characterize frequency combs, typically group index $n_\t{g}$ is replaced by effective refractive index $n_0$ in this formula, and often dispersion is ignored (thus $Y_0^\t{th}=1.54$ was used to infer $n_2$ of AlGaAs \cite{pu2016efficient_B} and Diamond \cite{hausmann2014diamond_B}).  
For weakly guiding waveguides made of low dispersion material, like Diamond \cite{hausmann2014diamond_B} or Si$_2$N$_3$ \cite{brasch2016photonic_B}, the discrepancy between $n_\t{g}$ and $n_0$ is negligible ($n_\t{g}\approx n_0$). In our case, however, $n_\t{g}/n_0>1.5$. We therefore adopt Eq. \ref{eq_Pth} explicitly (using $n_\t{g}$), to be consistent with \eqref{eqn_cavityeqofmotion} and \eqref{eq_dfdP}.  A similar approach seems prudent for AlGaAs \cite{pu2016efficient_B}. 

\begin{figure*}[t!]
	\includegraphics[width=2\columnwidth]{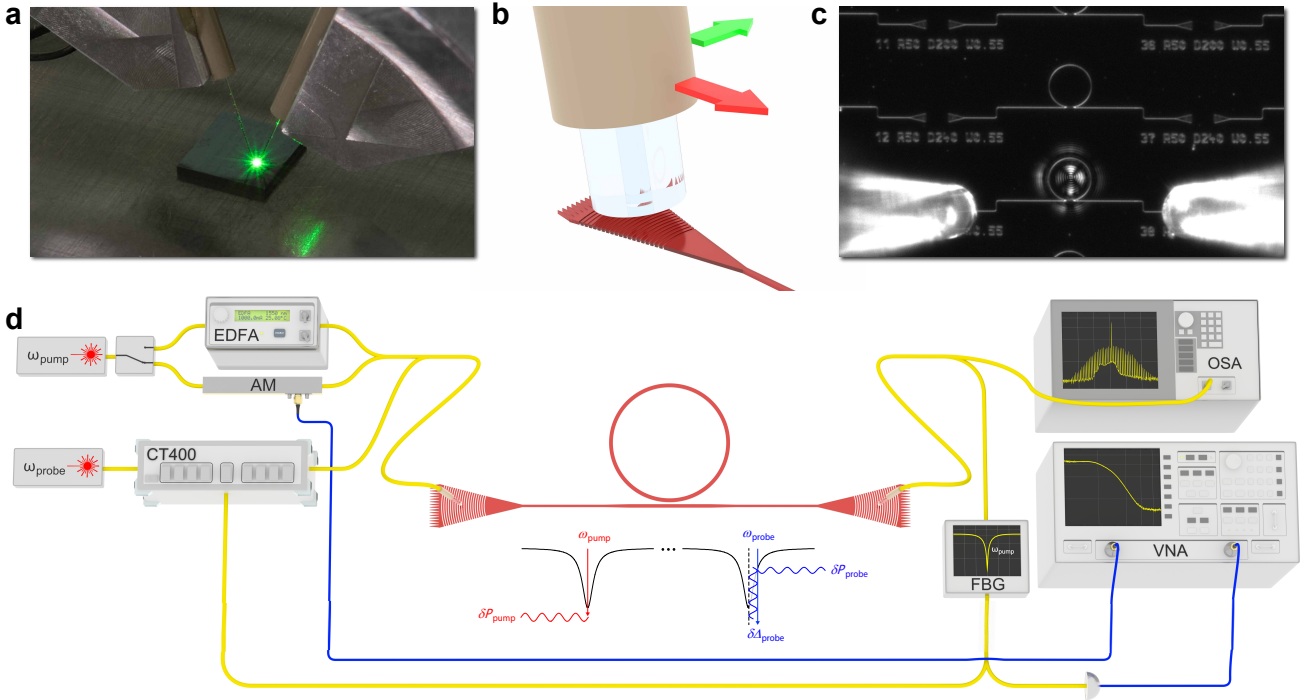}
	\caption{\textbf{Experimental setup.} (A) Photograph of a fiber-coupled ring resonator driven by $\sim100$ mW of 1550 nm light.  Green light is produced by 3rd-harmonic generation.  (B) Rendering of an optical fiber aligned to a waveguide grating coupler.  In practice the straight-cleaved facet of a SMF-28 fiber is positioned 10-20 $\mu$m above the chip at an angle 10-20 degrees from normal.  (C) Optical microscope image of a fiber-coupled ring resonator.  The ring is 50 $\mu$m in diameter.  Near-infrared light radiating from the ring is produced by 2nd-harmonic generation.  (D) Detailed schematic of the experiment.  Pump(probe) laser = Photonetics Tunics Plus tunable diode laser (10 mW, 1500-1600 nm).  EDFA = BkTel Erbium-doped fiber amplifier (34 dBm maximum output, 1535-1565 nm).  CT400 = Yenista CT400 Optical Component Tester.  IM = fiber electro-optic intensity modulator (10 GHz bandwidth). OSA = Optical spectrum analyzer (Agilent 86146  for near-infrared wavelengths and Yokogawa AQ6373 for visible wavelengths). FBG = Fiber-Bragg-grating-based optical notch filter (Advanced Optics Solutions, 1 nm bandwidth, 40 dB suppression, 1555-1565 nm). VNA = Agilent 8532 vector network analyzer (DC-500 MHz)}\label{fig_3}
\end{figure*}

\section{Measurement details}

Here we elaborate on the response measurements shown in Fig. 3 of the main text. A detailed schematic of the experiment is given in \fref{fig_3}.  As shown, the resonator is probed by two fields (``pump" and ``probe") generated by separate lasers.  The fields are combined before the resonator on a fiber beam splitter.  Before combining, the pump field is passed through one of three elements: a fast wavelength meter (CT400, for scanning transmission measurements), an intensity modulator (IM, for linear response measurements), or an optical amplifier (for frequency comb measurements).  At the resonator output, the combined field is split into two paths.  On one path, the output field is directed to an optical spectrum analyzer (OSA) for frequency comb measurements.  On the other path, the pump field is stripped using a fiber Bragg grating (FPG) and the filtered probe field is monitored by a high-speed photodetector for linear response measurements. A fraction of the field is also sent back to the CT400.  For linear response measurements, the IM is driven by the output of a vector network analyzer (VNA) and the input the VNA is connected to the output of the probe photodetector.

\subsection{Linear response measurement}

\begin{figure}[t!]
	\includegraphics[width=1.0\columnwidth]{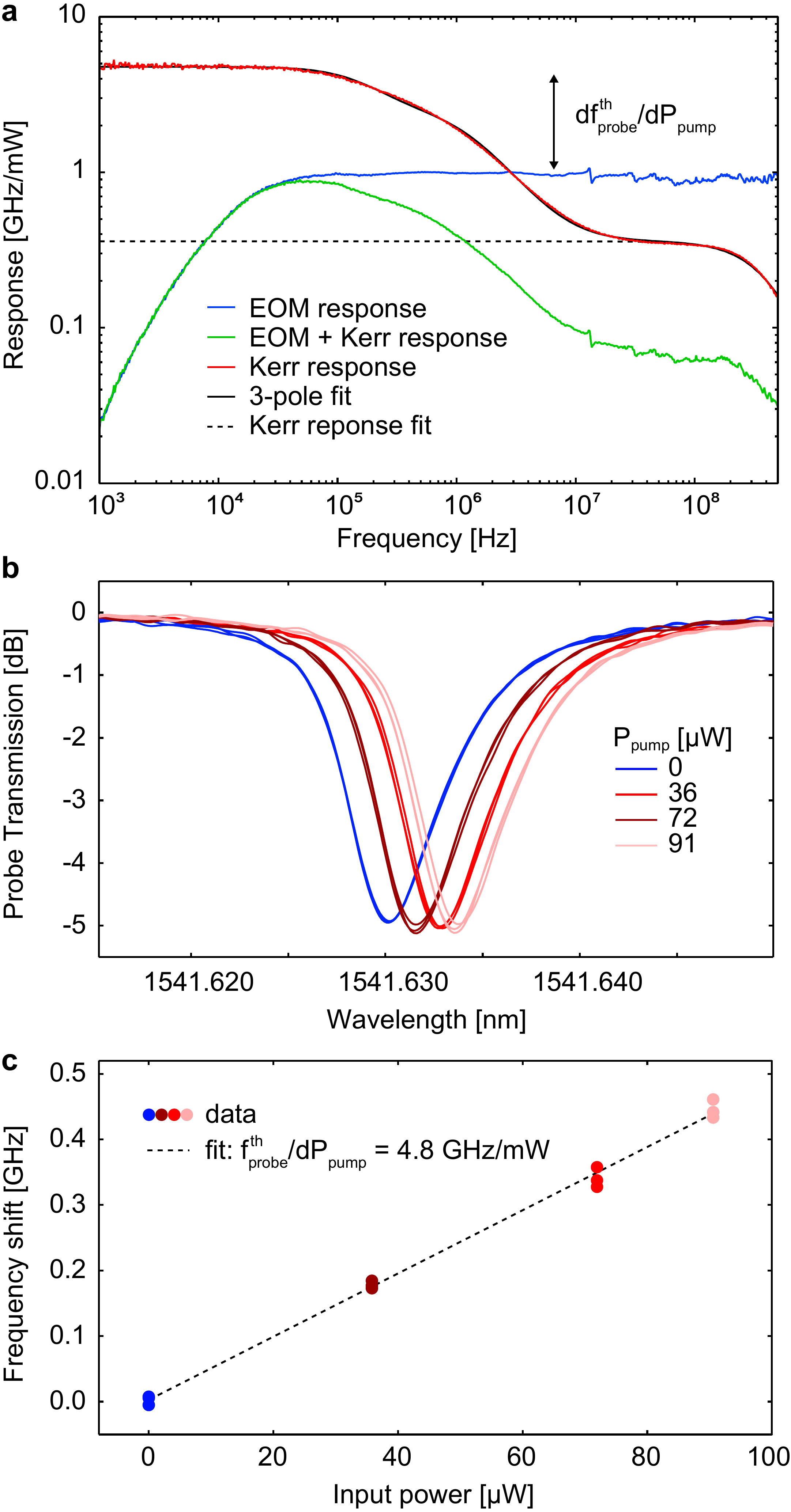}
	\caption{\textbf{Pump-probe response measurement details.} (a) Raw (blue, green) and processed (red) response measurements. (b) Probe transmission profile for various pump powers. (c) Linear fit to measurements of probe resonance frequency vs. pump power, used to bootstrap red curve in (a). }
	\label{fig_4}
\end{figure}

The linear microresonator response approach to measuring $n_2$ was first experimentally demonstrated, to our knowledge, in \cite{rokhsari2005observation_B}. The basic idea is to measure the resonance frequency shift produced by a weak modulation of the input power, using sufficiently low average power that the steady-state shift is small $(\ll\kappa)$.  In this linear regime, the response is given by combining Eqs. \ref{eq_dfdP}-\ref{eq_dfdP_xpm} with the cold form of \eqref{eqn_steadystateHOT} ($\Delta = \Delta_0$).  For resonant pumping ($\Delta_0 = 0$) using \eqref{eqn_steadystateHOTcrit} gives Eq. 1 of the main text,
\begin{subequations}\label{eq_dfdP_full}
	\begin{align}
	\frac{d\omega_{m'}}{d P_\t{in}^{(m)}}&\approx(2-\delta_{m,m'})\frac{2\eta\mathcal{F}_m}{\pi}\frac{\omega_m n_2}{n_\t{g} A_\t{eff}}\\
	&=(2-\delta_{m,m'})\frac{4\eta Q_m c}{V^{(m)}_\t{eff}}\frac{n_2}{n_\t{g}^2},
	\end{align}
\end{subequations}
where $m$ and $m'$ are the index of the pump and probe mode, respectively, and $\delta_{m,m'}$ is the Kronecker delta function.  The prefactor ensures that for distinct pump and probe modes $(m'\ne m)$, the shift produced by cross-phase-modulation is twice that produced by self-phase-modulation.

In practice, following \cite{rokhsari2005observation_B}, we use distinct pump and probe modes in order to minimize cross-talk. For the measurement shown in Fig. 3b of the main text, the pump field is tuned into resonance with a mode at 1560 nm and the probe field is detuned by $\Delta_0\approx -\kappa/2$ from an auxiliary mode at 1543 nm ($m-m' = 8$).  Nonzero probe detuning results in a output power modulation which is proportional to the resonance frequency modulation
\begin{equation}\label{eq_dPdf}
\delta P_\t{out}^{(m')}(\Omega) \approx 4\eta^2\frac{\partial\mathcal{L}}{\partial\Delta}\bar{P}_\t{out}^{(m')} \delta \omega_{m'}(\Omega)
\end{equation}
where $\Omega$ is the modulation frequency.

In principle it is possible to calibrate $\delta\omega_{m'}$  using \eqref{eq_dPdf} in conjunction with independent measurements of $\eta$, $\Delta_0$, $Q_{m'}$, $\bar{P}_\t{out}^{(m')}$, and $\delta\bar{P}_\t{out}^{(m')}$. This method was used in \cite{rokhsari2005observation_B}.  An alternative, more direct approach would be to separately modulate the probe field frequency by a known depth \cite{gorodetksy2010determination_B}.  To calibrate the response curve in Fig. 3b of the main text, we made use of a third, complementary approach that takes advantage of the fact that at low modulation frequencies, $\Omega\lesssim 2\pi\cdot 100\,\t{kHz}$, $\delta\omega_{m'}(\Omega)$ is dominated by a relatively large photothermal shift.  Namely, by measuring the static ($\Omega=0$) photothermal frequency shift of the probe resonance produced by the pump field, we calibrate the response curve according to:
%
\begin{equation}\label{eq_dfdP_calibration}
\frac{\delta\omega_{m'}(\Omega)}{\delta P_\t{in}^{(m)}(\Omega)}=\frac{\delta\omega_{m'}(0)}{\delta P_\t{in}^{(m)}(0)}\frac{\delta P_\t{in}^{(m)}(0)}{\delta P_\t{in}^{(m)}(\Omega)}\frac{\delta P_\t{out}^{(m')}(\Omega)}{\delta P_\t{out}^{(m')}(0)}
\end{equation}

From left to right, the first two terms on the right-hand-side of \eqref{eq_dfdP_calibration}  correspond to the static photothermal frequency shift and the transfer function of the pump intensity modulator, respectively.  Both are straightforward to determine (see \fref{fig_3} and \fref{fig_4}). The left hand side of \eqref{eq_dfdP_calibration} can be directly compared to \eqref{eq_dfdP_full} for modulation frequencies in which the Kerr effect dominates ($1\,\t{MHz} \lesssim \Omega/2\pi \lesssim 1\,\t{GHz}$ in \fref{fig_4}).  This method was used to determine $n_2$ as reported in the main text. Importantly, for the response measurement, the power of the pump and the probe was reduced to less than $100\,\mu\t{W}$ so that the static photothermal and Kerr shift of each beam is small compared to $\kappa$. Otherwise photothermal self-locking suppresses the low frequency response.

Details of the linear response measurement shown in Fig. 3b of the main text are given in \fref{fig_4}.  Green and blue lines in \fref{fig_4}a correspond to the response of the resonator output field and IM output field, respectively, to a modulation of the pump laser power. The latter is measured by analyzing a pickoff of the IM output field, as shown in \fref{fig_3}d.  Subtracting these curves (in dB units) gives the resonator response curve (red curve in \fref{fig_4}a).  The low frequency tail of this curve is normalized to the static photothermal frequency shift.  The latter is determined by plotting the resonance frequency of the probe beam versus pump power (\fref{fig_4}b,c).
\subsection{Frequency comb threshold measurements}

Frequency comb threshold measurements, shown in Fig. 2 of the main text, were also used to estimate $n_2$, utilizing Eq. 2.  The reported value $n_2 = 1.2(5)\cdot10^{-17}\,\t{m}^2/\t{W}$ was obtained from the average and standard deviation of the 16 measurements shown in Fig. 3d.  For each measurement, the resonance extinction ratio was used to infer $\eta$, via \eqref{eqn_steadystateHOTcrit}b.  An effective area of $A_\t{eff}\approx 0.2\,\mu\t{m}^2$ was assumed based on finite element simulation (\fref{fig_1}), this value being a good approximation for TE$_{00}$ modes of 500 and 550-nm-wide waveguides (both used in the measurement set).  A group index of $n_g=3.8$ was used based on measurements of the free spectral range (shown in Fig. 2h of main text). 

\subsection{Frequency comb noise measurements}

\begin{figure}[t!]
	\includegraphics[width=1.0\columnwidth]{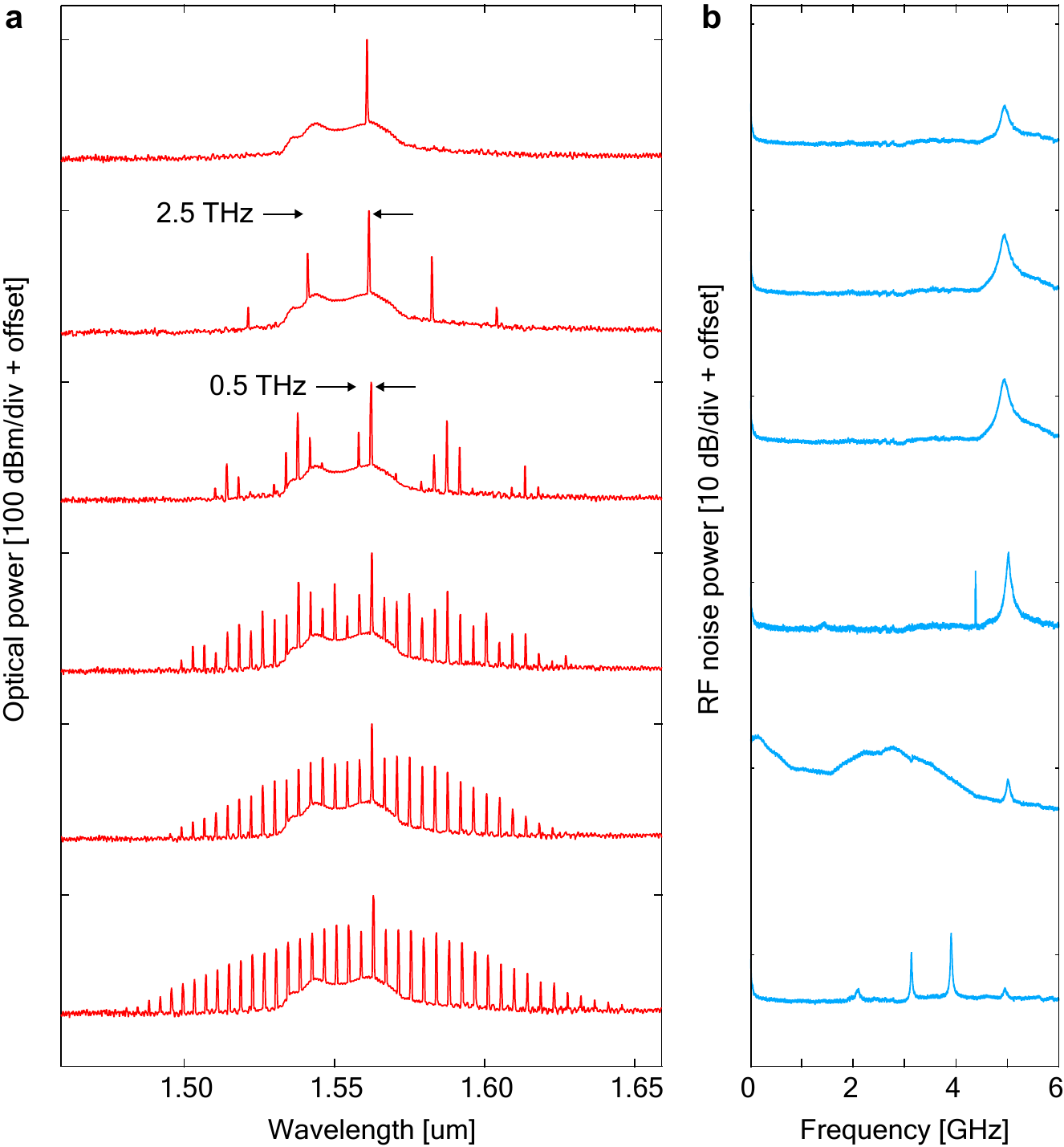}
	\caption{\textbf{Radiofrequency noise spectrum of Gap-on-insulator frequency comb.}  (a) Optical power spectrum of the output field of a $R=25\,\mu$m ring resonator pumped by a $\sim1$ W input field at 1.56 $\mu$m.  Sub-combs form and merge as laser-resonator detuning is decreased, as in Fig. 4 of main text. (b) Radiofrequency noise spectrum of same field, recorded by direct detection on a fast photodiode (JDSU RX10) and analyzed with an digital spectrum analyzer (Signal Hound). Elevated noise occurring as subcombs merge is evidence that the comb state is not coherent.}
	\label{fig_5}
\end{figure}

The sech-squared-shaped frequency comb shown in Fig. 4a of the main text --- observed for multiple devices --- is suggestive of a coherent (soliton) comb state. To test this hypothesis, the comb light was analyzed for radiofrequency (RF) intensity noise, its absence being a necessary (but not sufficient) condition for soliton formation.  To measure RF intensity noise, the comb light was directed to a fast photodetector and the photosignal was analyzed with a spectrum analyzer  (Signal Hound BB60C). Evolution of the RF noise spectrum of a 0.5 THz comb is shown in \fref{fig_5}. As the sub-combs merge, a large noise excess appears below 6 GHz (note that the persistent 5 GHz peak is intrinsic to the laser).  Similar behavior was observed in multiple devices.  We take this as evidence that the combs  observed are not fully coherent \cite{herr2012universal_B}.

%

\section{Fabrication Details}

\begin{figure}[b!]
	\includegraphics[width=1.0\columnwidth]{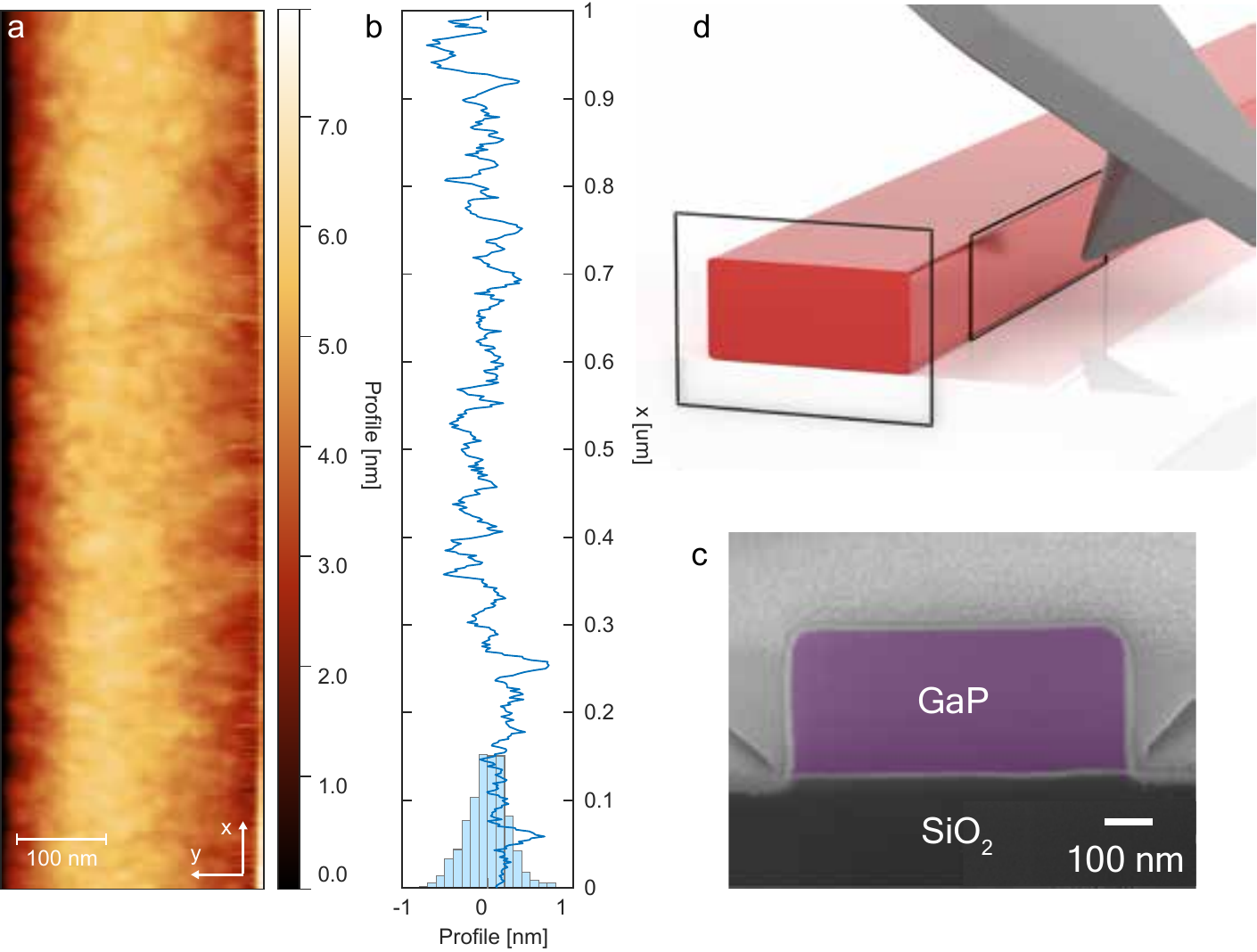}
	\caption{\textbf{Imaging of waveguide sidewall roughness and verticality.}  (a) AFM image of the sidewall of a 300-nm-thick waveguide. Color coding indicates displacement of the AFM tip versus positon on the sidewall.  (b) Vertical cut of the image in (a).  A histogram (bottom) gives a linear roughness of $\sim0.5$ nm, whereas the total (2D) roughness is 1.1 nm. (c) SEM image of a waveguide in cross-section, exposed by focused-ion-beam milling (same as Fig. 2d in main text). (d) Rendering of the configuration for AFM and SEM imaging.}
	\label{fig_6}
\end{figure}

Here we provide additional details on the fabrication process shown in Fig. 2a of the main text. We also refer the reader to \cite{schneider2018gallium_B} and \cite{honl2018highly_B}.

\subsection{GaP-on-insulator substrate fabrication}
Starting with a single-side-polished, 2-inch, [100]-oriented GaP wafer, a GaP/Al$_x$Ga$_{1-x}$P/GaP heterostructure is epitaxially grown, by metal-organic chemical vapor deposition.  The initial 100-nm-thick homoepitaxial layer of GaP facilitates nucleation and growth of the subsequent Al$_x$Ga$_{1-x}$P ($x=0.36$) etch-stop layer (also 100 nm thick), which is needed for later separation of the top GaP device layer (300 nm). All layers were deposited at a susceptor temperature of 650 $^\circ$C.  The precursors were trimethylgallium, trimethylaluminum and tertiarybutylphosphine.

In preparation for bonding, a thin film of Al$_2$O$_3$ is deposited by atomic layer deposition (ALD) on both the GaP device layer and a 4-inch silicon target wafer capped with 2 $\mu$m of SiO$_2$ prepared by thermal dry oxidation. The GaP wafer is bonded face down onto the oxidized silicon wafer, and the stack is annealed at 400 $^\circ$C to promote chemical bonding between the alumina layers. The majority of the initially 400-$\mu$m thick GaP substrate wafer is subsequently removed by wet etching in a solution of potassium ferricyanide (K$_3$Fe(CN)$_6$) and KOH. The remaining 50-100 $\mu$m of the substrate wafer are eliminated by inductively-coupled-plasma reactive ion etching (ICP-RIE) with a mixture of SiCl$_4$ and SF$_6$. The ICP-RIE process exhibits ultra-high selectivity for etching of GaP over Al$_x$Ga$_{1-x}$P, exceeding 1000:1 for $x = 0.36$, while maintaining an etch rate of 3 $\mu$m/min \cite{honl2018highly_B}. Finally, the Al$_x$Ga$_{1-x}$P layer is removed by submersion in concentrated HCl for 90 s, yielding the desired GaP-on-SiO$_2$ substrate. 

\subsection{Device fabrication}
All structures (waveguides, ring resonators, and grating couplers) are patterned in a single step by electron-beam lithography using 6$\%$  hydrogen silsesquioxane (HSQ) as resist (90 nm nominal thickness) and transferred into the GaP device layer by ICP-RIE using a Cl$_2$/BCl$_3$/H$_2$/CH$_4$ gas mixture. To promote adhesion of the HSQ, the surface of the GaP is coated by ALD with 3 nm of SiO$_2$ prior to spin-coating. After etching, the resist is stripped with buffered HF and a thin layer (5–10 nm) of Al$_2$O$_3$ is deposited by ALD onto the patterned chip to mitigate photo-oxidation. A subset of samples (Fig. 2f and Fig. 4c,d of main text) were additionally cladded with several  microns of SiO$_2$. The cladding is deposited conformally by plasma-enhanced chemical vapor deposition using tetraethylorthosilicate.

\subsection{Waveguide sidewalls}
Roughness and verticality of waveguide sidewalls were characterized by atomic force microscopy (AFM) and scanning electron microscopy (SEM), respectively.  Details are given in \fref{fig_6}. After deposition of the protective Al$_2$O$_3$ coating, an RMS roughness as low as 1.1 nm was observed (Fig. S4a). SEM of a waveguide cross-section exposed by focused ion beam milling (\fref{fig_6}b and Fig. 2d in the main text) reveals that the sidewalls are nearly vertical.


\begin{thebibliography}{45}%
	\makeatletter
	\providecommand \@ifxundefined [1]{%
		\@ifx{#1\undefined}
	}%
	\providecommand \@ifnum [1]{%
		\ifnum #1\expandafter \@firstoftwo
		\else \expandafter \@secondoftwo
		\fi
	}%
	\providecommand \@ifx [1]{%
		\ifx #1\expandafter \@firstoftwo
		\else \expandafter \@secondoftwo
		\fi
	}%
	\providecommand \natexlab [1]{#1}%
	\providecommand \enquote  [1]{``#1''}%
	\providecommand \bibnamefont  [1]{#1}%
	\providecommand \bibfnamefont [1]{#1}%
	\providecommand \citenamefont [1]{#1}%
	\providecommand \href@noop [0]{\@secondoftwo}%
	\providecommand \href [0]{\begingroup \@sanitize@url \@href}%
	\providecommand \@href[1]{\@@startlink{#1}\@@href}%
	\providecommand \@@href[1]{\endgroup#1\@@endlink}%
	\providecommand \@sanitize@url [0]{\catcode `\\12\catcode `\$12\catcode
		`\&12\catcode `\#12\catcode `\^12\catcode `\_12\catcode `\%12\relax}%
	\providecommand \@@startlink[1]{}%
	\providecommand \@@endlink[0]{}%
	\providecommand \url  [0]{\begingroup\@sanitize@url \@url }%
	\providecommand \@url [1]{\endgroup\@href {#1}{\urlprefix }}%
	\providecommand \urlprefix  [0]{URL }%
	\providecommand \Eprint [0]{\href }%
	\providecommand \doibase [0]{http://dx.doi.org/}%
	\providecommand \selectlanguage [0]{\@gobble}%
	\providecommand \bibinfo  [0]{\@secondoftwo}%
	\providecommand \bibfield  [0]{\@secondoftwo}%
	\providecommand \translation [1]{[#1]}%
	\providecommand \BibitemOpen [0]{}%
	\providecommand \bibitemStop [0]{}%
	\providecommand \bibitemNoStop [0]{.\EOS\space}%
	\providecommand \EOS [0]{\spacefactor3000\relax}%
	\providecommand \BibitemShut  [1]{\csname bibitem#1\endcsname}%
	\let\auto@bib@innerbib\@empty
	\bibitem [{\citenamefont {Pilkuhn}\ and\ \citenamefont
		{Foster}(1966)}]{pilkuhn1966green}%
	\BibitemOpen
	\bibfield  {author} {\bibinfo {author} {\bibfnamefont {M.}~\bibnamefont
			{Pilkuhn}}\ and\ \bibinfo {author} {\bibfnamefont {L.}~\bibnamefont
			{Foster}},\ }\href {http://ieeexplore.ieee.org/abstract/document/5392060/}
	{\bibfield  {journal} {\bibinfo  {journal} {IBM Journal of Research and
				Development}\ }\textbf {\bibinfo {volume} {10}},\ \bibinfo {pages} {122}
		(\bibinfo {year} {1966})}\BibitemShut {NoStop}%
	\bibitem [{\citenamefont {Mori}\ \emph {et~al.}(1987)\citenamefont {Mori},
		\citenamefont {Ogasawara}, \citenamefont {Yamamoto},\ and\ \citenamefont
		{Tachikawa}}]{mori1987new}%
	\BibitemOpen
	\bibfield  {author} {\bibinfo {author} {\bibfnamefont {H.}~\bibnamefont
			{Mori}}, \bibinfo {author} {\bibfnamefont {M.}~\bibnamefont {Ogasawara}},
		\bibinfo {author} {\bibfnamefont {M.}~\bibnamefont {Yamamoto}}, \ and\
		\bibinfo {author} {\bibfnamefont {M.}~\bibnamefont {Tachikawa}},\ }\href
	{https://aip.scitation.org/doi/abs/10.1063/1.98693} {\bibfield  {journal}
		{\bibinfo  {journal} {Applied physics letters}\ }\textbf {\bibinfo {volume}
			{51}},\ \bibinfo {pages} {1245} (\bibinfo {year} {1987})}\BibitemShut
	{NoStop}%
	\bibitem [{\citenamefont {Rivoire}\ \emph {et~al.}(2009)\citenamefont
		{Rivoire}, \citenamefont {Lin}, \citenamefont {Hatami}, \citenamefont
		{Masselink},\ and\ \citenamefont {Vu{\v{c}}kovi{\'c}}}]{rivoire2009second}%
	\BibitemOpen
	\bibfield  {author} {\bibinfo {author} {\bibfnamefont {K.}~\bibnamefont
			{Rivoire}}, \bibinfo {author} {\bibfnamefont {Z.}~\bibnamefont {Lin}},
		\bibinfo {author} {\bibfnamefont {F.}~\bibnamefont {Hatami}}, \bibinfo
		{author} {\bibfnamefont {W.~T.}\ \bibnamefont {Masselink}}, \ and\ \bibinfo
		{author} {\bibfnamefont {J.}~\bibnamefont {Vu{\v{c}}kovi{\'c}}},\ }\href
	{https://www.osapublishing.org/abstract.cfm?uri=oe-17-25-22609} {\bibfield
		{journal} {\bibinfo  {journal} {Opt. Exp.}\ }\textbf {\bibinfo {volume}
			{17}},\ \bibinfo {pages} {22609} (\bibinfo {year} {2009})}\BibitemShut
	{NoStop}%
	\bibitem [{\citenamefont {Lake}\ \emph {et~al.}(2016)\citenamefont {Lake},
		\citenamefont {Mitchell}, \citenamefont {Jayakumar}, \citenamefont {dos
			Santos}, \citenamefont {Curic},\ and\ \citenamefont
		{Barclay}}]{lake2016efficient}%
	\BibitemOpen
	\bibfield  {author} {\bibinfo {author} {\bibfnamefont {D.~P.}\ \bibnamefont
			{Lake}}, \bibinfo {author} {\bibfnamefont {M.}~\bibnamefont {Mitchell}},
		\bibinfo {author} {\bibfnamefont {H.}~\bibnamefont {Jayakumar}}, \bibinfo
		{author} {\bibfnamefont {L.~F.}\ \bibnamefont {dos Santos}}, \bibinfo
		{author} {\bibfnamefont {D.}~\bibnamefont {Curic}}, \ and\ \bibinfo {author}
		{\bibfnamefont {P.~E.}\ \bibnamefont {Barclay}},\ }\href
	{http://aip.scitation.org/doi/abs/10.1063/1.4940242} {\bibfield  {journal}
		{\bibinfo  {journal} {Appl. Phys. Lett.}\ }\textbf {\bibinfo {volume}
			{108}},\ \bibinfo {pages} {031109} (\bibinfo {year} {2016})}\BibitemShut
	{NoStop}%
	\bibitem [{\citenamefont {Gan}\ \emph {et~al.}(2012)\citenamefont {Gan},
		\citenamefont {Pervez}, \citenamefont {Kymissis}, \citenamefont {Hatami},\
		and\ \citenamefont {Englund}}]{gan2012high}%
	\BibitemOpen
	\bibfield  {author} {\bibinfo {author} {\bibfnamefont {X.}~\bibnamefont
			{Gan}}, \bibinfo {author} {\bibfnamefont {N.}~\bibnamefont {Pervez}},
		\bibinfo {author} {\bibfnamefont {I.}~\bibnamefont {Kymissis}}, \bibinfo
		{author} {\bibfnamefont {F.}~\bibnamefont {Hatami}}, \ and\ \bibinfo {author}
		{\bibfnamefont {D.}~\bibnamefont {Englund}},\ }\href {\doibase
		10.1063/1.4724177} {\bibfield  {journal} {\bibinfo  {journal} {Appl. Phys.
				Lett.}\ }\textbf {\bibinfo {volume} {100}},\ \bibinfo {pages} {231104}
		(\bibinfo {year} {2012})}\BibitemShut {NoStop}%
	\bibitem [{\citenamefont {Gonz{\'a}lez-Tudela}\ \emph
		{et~al.}(2015)\citenamefont {Gonz{\'a}lez-Tudela}, \citenamefont {Hung},
		\citenamefont {Chang}, \citenamefont {Cirac},\ and\ \citenamefont
		{Kimble}}]{gonzalez2015subwavelength}%
	\BibitemOpen
	\bibfield  {author} {\bibinfo {author} {\bibfnamefont {A.}~\bibnamefont
			{Gonz{\'a}lez-Tudela}}, \bibinfo {author} {\bibfnamefont {C.-L.}\
			\bibnamefont {Hung}}, \bibinfo {author} {\bibfnamefont {D.~E.}\ \bibnamefont
			{Chang}}, \bibinfo {author} {\bibfnamefont {J.~I.}\ \bibnamefont {Cirac}}, \
		and\ \bibinfo {author} {\bibfnamefont {H.}~\bibnamefont {Kimble}},\ }\href
	{https://www.nature.com/nphoton/journal/v9/n5/abs/nphoton.2015.54.html}
	{\bibfield  {journal} {\bibinfo  {journal} {Nature Photonics}\ }\textbf
		{\bibinfo {volume} {9}},\ \bibinfo {pages} {320} (\bibinfo {year}
		{2015})}\BibitemShut {NoStop}%
	\bibitem [{\citenamefont {Englund}\ \emph {et~al.}(2010)\citenamefont
		{Englund}, \citenamefont {Shields}, \citenamefont {Rivoire}, \citenamefont
		{Hatami}, \citenamefont {Vuckovic}, \citenamefont {Park},\ and\ \citenamefont
		{Lukin}}]{englund2010deterministic}%
	\BibitemOpen
	\bibfield  {author} {\bibinfo {author} {\bibfnamefont {D.}~\bibnamefont
			{Englund}}, \bibinfo {author} {\bibfnamefont {B.}~\bibnamefont {Shields}},
		\bibinfo {author} {\bibfnamefont {K.}~\bibnamefont {Rivoire}}, \bibinfo
		{author} {\bibfnamefont {F.}~\bibnamefont {Hatami}}, \bibinfo {author}
		{\bibfnamefont {J.}~\bibnamefont {Vuckovic}}, \bibinfo {author}
		{\bibfnamefont {H.}~\bibnamefont {Park}}, \ and\ \bibinfo {author}
		{\bibfnamefont {M.~D.}\ \bibnamefont {Lukin}},\ }\href
	{https://pubs.acs.org/doi/abs/10.1021/nl101662v} {\bibfield  {journal}
		{\bibinfo  {journal} {Nano Lett.}\ }\textbf {\bibinfo {volume} {10}},\
		\bibinfo {pages} {3922} (\bibinfo {year} {2010})}\BibitemShut {NoStop}%
	\bibitem [{\citenamefont {Gould}\ \emph {et~al.}(2016)\citenamefont {Gould},
		\citenamefont {Schmidgall}, \citenamefont {Dadgostar}, \citenamefont
		{Hatami},\ and\ \citenamefont {Fu}}]{gould2016efficient}%
	\BibitemOpen
	\bibfield  {author} {\bibinfo {author} {\bibfnamefont {M.}~\bibnamefont
			{Gould}}, \bibinfo {author} {\bibfnamefont {E.~R.}\ \bibnamefont
			{Schmidgall}}, \bibinfo {author} {\bibfnamefont {S.}~\bibnamefont
			{Dadgostar}}, \bibinfo {author} {\bibfnamefont {F.}~\bibnamefont {Hatami}}, \
		and\ \bibinfo {author} {\bibfnamefont {K.-M.~C.}\ \bibnamefont {Fu}},\ }\href
	{\doibase 10.1103/PhysRevApplied.6.011001} {\bibfield  {journal} {\bibinfo
			{journal} {Phys. Rev. Appl.}\ }\textbf {\bibinfo {volume} {6}},\ \bibinfo
		{pages} {011001} (\bibinfo {year} {2016})}\BibitemShut {NoStop}%
	\bibitem [{\citenamefont {Schneider}\ \emph {et~al.}(2018)\citenamefont
		{Schneider}, \citenamefont {Welter}, \citenamefont {Baumgartner},
		\citenamefont {Hahn}, \citenamefont {Czornomaz},\ and\ \citenamefont
		{Seidler}}]{schneider2018gallium}%
	\BibitemOpen
	\bibfield  {author} {\bibinfo {author} {\bibfnamefont {K.}~\bibnamefont
			{Schneider}}, \bibinfo {author} {\bibfnamefont {P.}~\bibnamefont {Welter}},
		\bibinfo {author} {\bibfnamefont {Y.}~\bibnamefont {Baumgartner}}, \bibinfo
		{author} {\bibfnamefont {H.}~\bibnamefont {Hahn}}, \bibinfo {author}
		{\bibfnamefont {L.}~\bibnamefont {Czornomaz}}, \ and\ \bibinfo {author}
		{\bibfnamefont {P.}~\bibnamefont {Seidler}},\ }\href
	{https://www.osapublishing.org/abstract.cfm?uri=jlt-36-14-2994} {\bibfield
		{journal} {\bibinfo  {journal} {Journal of Lightwave Technology}\ }\textbf
		{\bibinfo {volume} {36}},\ \bibinfo {pages} {2994} (\bibinfo {year}
		{2018})}\BibitemShut {NoStop}%
	\bibitem [{\citenamefont {P}\ \emph {et~al.}(1999)\citenamefont {P},
		\citenamefont {Schuurmans}, \citenamefont {Vanmaekelbergh}, \citenamefont
		{Lagemaat},\ and\ \citenamefont {Lagendijk}}]{p1999strongly}%
	\BibitemOpen
	\bibfield  {author} {\bibinfo {author} {\bibfnamefont {F.~J.}\ \bibnamefont
			{P}}, \bibinfo {author} {\bibnamefont {Schuurmans}}, \bibinfo {author}
		{\bibfnamefont {D.}~\bibnamefont {Vanmaekelbergh}}, \bibinfo {author}
		{\bibfnamefont {J.~v.~d.}\ \bibnamefont {Lagemaat}}, \ and\ \bibinfo {author}
		{\bibfnamefont {A.}~\bibnamefont {Lagendijk}},\ }\href {\doibase
		10.1126/science.284.5411.141} {\bibfield  {journal} {\bibinfo  {journal}
			{Science}\ }\textbf {\bibinfo {volume} {284}},\ \bibinfo {pages} {141}
		(\bibinfo {year} {1999})}\BibitemShut {NoStop}%
	\bibitem [{\citenamefont {Barclay}\ \emph {et~al.}(2009)\citenamefont
		{Barclay}, \citenamefont {Fu}, \citenamefont {Santori},\ and\ \citenamefont
		{Beausoleil}}]{barclay2009chip}%
	\BibitemOpen
	\bibfield  {author} {\bibinfo {author} {\bibfnamefont {P.~E.}\ \bibnamefont
			{Barclay}}, \bibinfo {author} {\bibfnamefont {K.-M.~C.}\ \bibnamefont {Fu}},
		\bibinfo {author} {\bibfnamefont {C.}~\bibnamefont {Santori}}, \ and\
		\bibinfo {author} {\bibfnamefont {R.~G.}\ \bibnamefont {Beausoleil}},\ }\href
	{\doibase 10.1063/1.3262948} {\bibfield  {journal} {\bibinfo  {journal}
			{Appl. Phys. Lett.}\ }\textbf {\bibinfo {volume} {95}},\ \bibinfo {pages}
		{191115} (\bibinfo {year} {2009})}\BibitemShut {NoStop}%
	\bibitem [{\citenamefont {Mitchell}\ \emph {et~al.}(2014)\citenamefont
		{Mitchell}, \citenamefont {Hryciw},\ and\ \citenamefont
		{Barclay}}]{mitchell2014cavity}%
	\BibitemOpen
	\bibfield  {author} {\bibinfo {author} {\bibfnamefont {M.}~\bibnamefont
			{Mitchell}}, \bibinfo {author} {\bibfnamefont {A.~C.}\ \bibnamefont
			{Hryciw}}, \ and\ \bibinfo {author} {\bibfnamefont {P.~E.}\ \bibnamefont
			{Barclay}},\ }\href {http://aip.scitation.org/doi/abs/10.1063/1.4870999}
	{\bibfield  {journal} {\bibinfo  {journal} {Appl. Phys. Lett.}\ }\textbf
		{\bibinfo {volume} {104}},\ \bibinfo {pages} {141104} (\bibinfo {year}
		{2014})}\BibitemShut {NoStop}%
	\bibitem [{\citenamefont {Guillemé}\ \emph {et~al.}(2017)\citenamefont
		{Guillemé}, \citenamefont {Dumeige}, \citenamefont {Stodolna}, \citenamefont
		{Vallet}, \citenamefont {Rohel}, \citenamefont {Létoublon}, \citenamefont
		{Cornet}, \citenamefont {Ponchet}, \citenamefont {Durand},\ and\
		\citenamefont {Léger}}]{guilleme2017second}%
	\BibitemOpen
	\bibfield  {author} {\bibinfo {author} {\bibfnamefont {P.}~\bibnamefont
			{Guillemé}}, \bibinfo {author} {\bibfnamefont {Y.}~\bibnamefont {Dumeige}},
		\bibinfo {author} {\bibfnamefont {J.}~\bibnamefont {Stodolna}}, \bibinfo
		{author} {\bibfnamefont {M.}~\bibnamefont {Vallet}}, \bibinfo {author}
		{\bibfnamefont {T.}~\bibnamefont {Rohel}}, \bibinfo {author} {\bibfnamefont
			{A.}~\bibnamefont {Létoublon}}, \bibinfo {author} {\bibfnamefont
			{C.}~\bibnamefont {Cornet}}, \bibinfo {author} {\bibfnamefont
			{A.}~\bibnamefont {Ponchet}}, \bibinfo {author} {\bibfnamefont
			{O.}~\bibnamefont {Durand}}, \ and\ \bibinfo {author} {\bibfnamefont
			{Y.}~\bibnamefont {Léger}},\ }\href {\doibase 10.1088/1361-6641/aa676d}
	{\bibfield  {journal} {\bibinfo  {journal} {Semicond. Sci. Technol.}\
		}\textbf {\bibinfo {volume} {32}},\ \bibinfo {pages} {065004} (\bibinfo
		{year} {2017})}\BibitemShut {NoStop}%
	\bibitem [{\citenamefont {Thomas}\ \emph {et~al.}(2014)\citenamefont {Thomas},
		\citenamefont {Barbour}, \citenamefont {Song}, \citenamefont {Lee},\ and\
		\citenamefont {Fu}}]{thomas2014waveguide}%
	\BibitemOpen
	\bibfield  {author} {\bibinfo {author} {\bibfnamefont {N.}~\bibnamefont
			{Thomas}}, \bibinfo {author} {\bibfnamefont {R.~J.}\ \bibnamefont {Barbour}},
		\bibinfo {author} {\bibfnamefont {Y.}~\bibnamefont {Song}}, \bibinfo {author}
		{\bibfnamefont {M.~L.}\ \bibnamefont {Lee}}, \ and\ \bibinfo {author}
		{\bibfnamefont {K.-M.~C.}\ \bibnamefont {Fu}},\ }\href {\doibase
		10.1364/OE.22.013555} {\bibfield  {journal} {\bibinfo  {journal} {Opt. Exp.}\
		}\textbf {\bibinfo {volume} {22}},\ \bibinfo {pages} {13555} (\bibinfo {year}
		{2014})}\BibitemShut {NoStop}%
	\bibitem [{\citenamefont {Cambiasso}\ \emph {et~al.}(2017)\citenamefont
		{Cambiasso}, \citenamefont {Grinblat}, \citenamefont {Li}, \citenamefont
		{Rakovich}, \citenamefont {Cortés},\ and\ \citenamefont
		{Maier}}]{cambiasso2017bridging}%
	\BibitemOpen
	\bibfield  {author} {\bibinfo {author} {\bibfnamefont {J.}~\bibnamefont
			{Cambiasso}}, \bibinfo {author} {\bibfnamefont {G.}~\bibnamefont {Grinblat}},
		\bibinfo {author} {\bibfnamefont {Y.}~\bibnamefont {Li}}, \bibinfo {author}
		{\bibfnamefont {A.}~\bibnamefont {Rakovich}}, \bibinfo {author}
		{\bibfnamefont {E.}~\bibnamefont {Cortés}}, \ and\ \bibinfo {author}
		{\bibfnamefont {S.~A.}\ \bibnamefont {Maier}},\ }\href {\doibase
		10.1021/acs.nanolett.6b05026} {\bibfield  {journal} {\bibinfo  {journal}
			{Nano Lett.}\ }\textbf {\bibinfo {volume} {17}},\ \bibinfo {pages} {1219}
		(\bibinfo {year} {2017})}\BibitemShut {NoStop}%
	\bibitem [{\citenamefont {Pu}\ \emph {et~al.}(2016)\citenamefont {Pu},
		\citenamefont {Ottaviano}, \citenamefont {Semenova},\ and\ \citenamefont
		{Yvind}}]{pu2016efficient}%
	\BibitemOpen
	\bibfield  {author} {\bibinfo {author} {\bibfnamefont {M.}~\bibnamefont
			{Pu}}, \bibinfo {author} {\bibfnamefont {L.}~\bibnamefont {Ottaviano}},
		\bibinfo {author} {\bibfnamefont {E.}~\bibnamefont {Semenova}}, \ and\
		\bibinfo {author} {\bibfnamefont {K.}~\bibnamefont {Yvind}},\ }\href
	{https://www.osapublishing.org/abstract.cfm?uri=optica-3-8-823} {\bibfield
		{journal} {\bibinfo  {journal} {Optica}\ }\textbf {\bibinfo {volume} {3}},\
		\bibinfo {pages} {823} (\bibinfo {year} {2016})}\BibitemShut {NoStop}%
	\bibitem [{\citenamefont {Lu}\ \emph {et~al.}(2014)\citenamefont {Lu},
		\citenamefont {Lee}, \citenamefont {Rogers},\ and\ \citenamefont
		{Lin}}]{lu2014optical}%
	\BibitemOpen
	\bibfield  {author} {\bibinfo {author} {\bibfnamefont {X.}~\bibnamefont
			{Lu}}, \bibinfo {author} {\bibfnamefont {J.~Y.}\ \bibnamefont {Lee}},
		\bibinfo {author} {\bibfnamefont {S.}~\bibnamefont {Rogers}}, \ and\ \bibinfo
		{author} {\bibfnamefont {Q.}~\bibnamefont {Lin}},\ }\href
	{https://www.osapublishing.org/abstract.cfm?uri=oe-22-25-30826} {\bibfield
		{journal} {\bibinfo  {journal} {Opt. Exp.}\ }\textbf {\bibinfo {volume}
			{22}},\ \bibinfo {pages} {30826} (\bibinfo {year} {2014})}\BibitemShut
	{NoStop}%
	\bibitem [{\citenamefont {Dinu}\ \emph {et~al.}(2003)\citenamefont {Dinu},
		\citenamefont {Quochi},\ and\ \citenamefont {Garcia}}]{dinu2003third}%
	\BibitemOpen
	\bibfield  {author} {\bibinfo {author} {\bibfnamefont {M.}~\bibnamefont
			{Dinu}}, \bibinfo {author} {\bibfnamefont {F.}~\bibnamefont {Quochi}}, \ and\
		\bibinfo {author} {\bibfnamefont {H.}~\bibnamefont {Garcia}},\ }\href
	{https://aip.scitation.org/doi/abs/10.1063/1.1571665} {\bibfield  {journal}
		{\bibinfo  {journal} {Appl. Phys. Lett.}\ }\textbf {\bibinfo {volume} {82}},\
		\bibinfo {pages} {2954} (\bibinfo {year} {2003})}\BibitemShut {NoStop}%
	\bibitem [{\citenamefont {Sun}\ \emph {et~al.}(2000)\citenamefont {Sun},
		\citenamefont {Chu}, \citenamefont {Keller},\ and\ \citenamefont
		{DenBaars}}]{sun2000third}%
	\BibitemOpen
	\bibfield  {author} {\bibinfo {author} {\bibfnamefont {C.-K.}\ \bibnamefont
			{Sun}}, \bibinfo {author} {\bibfnamefont {S.-W.}\ \bibnamefont {Chu}},
		\bibinfo {author} {\bibfnamefont {S.}~\bibnamefont {Keller}}, \ and\ \bibinfo
		{author} {\bibfnamefont {S.}~\bibnamefont {DenBaars}},\ }in\ \href
	{https://ieeexplore.ieee.org/abstract/document/907762/} {\emph {\bibinfo
			{booktitle} {Quantum Electronics Conference, 2000. Conference Digest. 2000
				International}}}\ (\bibinfo {organization} {IEEE},\ \bibinfo {year} {2000})\
	pp.\ \bibinfo {pages} {1--pp}\BibitemShut {NoStop}%
	\bibitem [{\citenamefont {Ching}\ and\ \citenamefont
		{Huang}(1993)}]{ching1993calculation}%
	\BibitemOpen
	\bibfield  {author} {\bibinfo {author} {\bibfnamefont {W.}~\bibnamefont
			{Ching}}\ and\ \bibinfo {author} {\bibfnamefont {M.-Z.}\ \bibnamefont
			{Huang}},\ }\href
	{https://journals.aps.org/prb/abstract/10.1103/PhysRevB.47.9479} {\bibfield
		{journal} {\bibinfo  {journal} {Phys. Rev. B}\ }\textbf {\bibinfo {volume}
			{47}},\ \bibinfo {pages} {9479} (\bibinfo {year} {1993})}\BibitemShut
	{NoStop}%
	\bibitem [{\citenamefont {Martin}\ \emph {et~al.}(2017)\citenamefont {Martin},
		\citenamefont {Sanchez}, \citenamefont {Combri{\'e}}, \citenamefont
		{De~Rossi},\ and\ \citenamefont {Raineri}}]{martin2017gainp}%
	\BibitemOpen
	\bibfield  {author} {\bibinfo {author} {\bibfnamefont {A.}~\bibnamefont
			{Martin}}, \bibinfo {author} {\bibfnamefont {D.}~\bibnamefont {Sanchez}},
		\bibinfo {author} {\bibfnamefont {S.}~\bibnamefont {Combri{\'e}}}, \bibinfo
		{author} {\bibfnamefont {A.}~\bibnamefont {De~Rossi}}, \ and\ \bibinfo
		{author} {\bibfnamefont {F.}~\bibnamefont {Raineri}},\ }\href
	{https://www.osapublishing.org/abstract.cfm?uri=ol-42-3-599} {\bibfield
		{journal} {\bibinfo  {journal} {Opt. Lett.}\ }\textbf {\bibinfo {volume}
			{42}},\ \bibinfo {pages} {599} (\bibinfo {year} {2017})}\BibitemShut
	{NoStop}%
	\bibitem [{\citenamefont {Hurlbut}\ \emph {et~al.}(2007)\citenamefont
		{Hurlbut}, \citenamefont {Lee}, \citenamefont {Vodopyanov}, \citenamefont
		{Kuo},\ and\ \citenamefont {Fejer}}]{hurlbut2007multiphoton}%
	\BibitemOpen
	\bibfield  {author} {\bibinfo {author} {\bibfnamefont {W.~C.}\ \bibnamefont
			{Hurlbut}}, \bibinfo {author} {\bibfnamefont {Y.-S.}\ \bibnamefont {Lee}},
		\bibinfo {author} {\bibfnamefont {K.}~\bibnamefont {Vodopyanov}}, \bibinfo
		{author} {\bibfnamefont {P.}~\bibnamefont {Kuo}}, \ and\ \bibinfo {author}
		{\bibfnamefont {M.}~\bibnamefont {Fejer}},\ }\href
	{https://www.osapublishing.org/abstract.cfm?uri=ol-32-6-668} {\bibfield
		{journal} {\bibinfo  {journal} {Opt. Lett.}\ }\textbf {\bibinfo {volume}
			{32}},\ \bibinfo {pages} {668} (\bibinfo {year} {2007})}\BibitemShut
	{NoStop}%
	\bibitem [{\citenamefont {Liu}\ \emph {et~al.}(2010)\citenamefont {Liu},
		\citenamefont {Li}, \citenamefont {Xing}, \citenamefont {Chai}, \citenamefont
		{Hu}, \citenamefont {Wang}, \citenamefont {Deng}, \citenamefont {Sun},\ and\
		\citenamefont {Wang}}]{liu2010three}%
	\BibitemOpen
	\bibfield  {author} {\bibinfo {author} {\bibfnamefont {F.}~\bibnamefont
			{Liu}}, \bibinfo {author} {\bibfnamefont {Y.}~\bibnamefont {Li}}, \bibinfo
		{author} {\bibfnamefont {Q.}~\bibnamefont {Xing}}, \bibinfo {author}
		{\bibfnamefont {L.}~\bibnamefont {Chai}}, \bibinfo {author} {\bibfnamefont
			{M.}~\bibnamefont {Hu}}, \bibinfo {author} {\bibfnamefont {C.}~\bibnamefont
			{Wang}}, \bibinfo {author} {\bibfnamefont {Y.}~\bibnamefont {Deng}}, \bibinfo
		{author} {\bibfnamefont {Q.}~\bibnamefont {Sun}}, \ and\ \bibinfo {author}
		{\bibfnamefont {C.}~\bibnamefont {Wang}},\ }\href
	{http://iopscience.iop.org/article/10.1088/2040-8978/12/9/095201/meta}
	{\bibfield  {journal} {\bibinfo  {journal} {Journal of Optics}\ }\textbf
		{\bibinfo {volume} {12}},\ \bibinfo {pages} {095201} (\bibinfo {year}
		{2010})}\BibitemShut {NoStop}%
	\bibitem [{\citenamefont {Martin}\ \emph {et~al.}(2018)\citenamefont {Martin},
		\citenamefont {Combri{\'e}}, \citenamefont {de~Rossi}, \citenamefont
		{Beaudoin}, \citenamefont {Sagnes},\ and\ \citenamefont
		{Raineri}}]{martin2018nonlinear}%
	\BibitemOpen
	\bibfield  {author} {\bibinfo {author} {\bibfnamefont {A.}~\bibnamefont
			{Martin}}, \bibinfo {author} {\bibfnamefont {S.}~\bibnamefont {Combri{\'e}}},
		\bibinfo {author} {\bibfnamefont {A.}~\bibnamefont {de~Rossi}}, \bibinfo
		{author} {\bibfnamefont {G.}~\bibnamefont {Beaudoin}}, \bibinfo {author}
		{\bibfnamefont {I.}~\bibnamefont {Sagnes}}, \ and\ \bibinfo {author}
		{\bibfnamefont {F.}~\bibnamefont {Raineri}},\ }\href
	{https://www.osapublishing.org/abstract.cfm?uri=prj-6-5-B43} {\bibfield
		{journal} {\bibinfo  {journal} {Phot. Res.}\ }\textbf {\bibinfo {volume}
			{6}},\ \bibinfo {pages} {B43} (\bibinfo {year} {2018})}\BibitemShut {NoStop}%
	\bibitem [{\citenamefont {Skryabin}\ and\ \citenamefont
		{Gorbach}(2010)}]{skryabin2010colloquium}%
	\BibitemOpen
	\bibfield  {author} {\bibinfo {author} {\bibfnamefont {D.~V.}\ \bibnamefont
			{Skryabin}}\ and\ \bibinfo {author} {\bibfnamefont {A.~V.}\ \bibnamefont
			{Gorbach}},\ }\href
	{https://journals.aps.org/rmp/abstract/10.1103/RevModPhys.82.1287} {\bibfield
		{journal} {\bibinfo  {journal} {Rev. Mod. Phys.}\ }\textbf {\bibinfo
			{volume} {82}},\ \bibinfo {pages} {1287} (\bibinfo {year}
		{2010})}\BibitemShut {NoStop}%
	\bibitem [{\citenamefont {Halir}\ \emph {et~al.}(2012)\citenamefont {Halir},
		\citenamefont {Okawachi}, \citenamefont {Levy}, \citenamefont {Foster},
		\citenamefont {Lipson},\ and\ \citenamefont
		{Gaeta}}]{halir2012ultrabroadband}%
	\BibitemOpen
	\bibfield  {author} {\bibinfo {author} {\bibfnamefont {R.}~\bibnamefont
			{Halir}}, \bibinfo {author} {\bibfnamefont {Y.}~\bibnamefont {Okawachi}},
		\bibinfo {author} {\bibfnamefont {J.}~\bibnamefont {Levy}}, \bibinfo {author}
		{\bibfnamefont {M.}~\bibnamefont {Foster}}, \bibinfo {author} {\bibfnamefont
			{M.}~\bibnamefont {Lipson}}, \ and\ \bibinfo {author} {\bibfnamefont
			{A.}~\bibnamefont {Gaeta}},\ }\href
	{https://www.osapublishing.org/abstract.cfm?uri=ol-37-10-1685} {\bibfield
		{journal} {\bibinfo  {journal} {Opt. Lett.}\ }\textbf {\bibinfo {volume}
			{37}},\ \bibinfo {pages} {1685} (\bibinfo {year} {2012})}\BibitemShut
	{NoStop}%
	\bibitem [{\citenamefont {Del’Haye}\ \emph {et~al.}(2007)\citenamefont
		{Del’Haye}, \citenamefont {Schliesser}, \citenamefont {Arcizet},
		\citenamefont {Wilken}, \citenamefont {Holzwarth},\ and\ \citenamefont
		{Kippenberg}}]{del2007optical}%
	\BibitemOpen
	\bibfield  {author} {\bibinfo {author} {\bibfnamefont {P.}~\bibnamefont
			{Del’Haye}}, \bibinfo {author} {\bibfnamefont {A.}~\bibnamefont
			{Schliesser}}, \bibinfo {author} {\bibfnamefont {O.}~\bibnamefont {Arcizet}},
		\bibinfo {author} {\bibfnamefont {T.}~\bibnamefont {Wilken}}, \bibinfo
		{author} {\bibfnamefont {R.}~\bibnamefont {Holzwarth}}, \ and\ \bibinfo
		{author} {\bibfnamefont {T.~J.}\ \bibnamefont {Kippenberg}},\ }\href
	{https://www.nature.com/articles/nature06401} {\bibfield  {journal} {\bibinfo
			{journal} {Nature}\ }\textbf {\bibinfo {volume} {450}},\ \bibinfo {pages}
		{1214} (\bibinfo {year} {2007})}\BibitemShut {NoStop}%
	\bibitem [{\citenamefont {Kippenberg}\ \emph {et~al.}(2011)\citenamefont
		{Kippenberg}, \citenamefont {Holzwarth},\ and\ \citenamefont
		{Diddams}}]{kippenberg2011microresonator}%
	\BibitemOpen
	\bibfield  {author} {\bibinfo {author} {\bibfnamefont {T.~J.}\ \bibnamefont
			{Kippenberg}}, \bibinfo {author} {\bibfnamefont {R.}~\bibnamefont
			{Holzwarth}}, \ and\ \bibinfo {author} {\bibfnamefont {S.~A.}\ \bibnamefont
			{Diddams}},\ }\href {http://science.sciencemag.org/content/332/6029/555}
	{\bibfield  {journal} {\bibinfo  {journal} {Science}\ }\textbf {\bibinfo
			{volume} {332}},\ \bibinfo {pages} {555} (\bibinfo {year}
		{2011})}\BibitemShut {NoStop}%
	\bibitem [{\citenamefont {Herr}\ \emph
		{et~al.}(2014{\natexlab{a}})\citenamefont {Herr}, \citenamefont {Brasch},
		\citenamefont {Jost}, \citenamefont {Wang}, \citenamefont {Kondratiev},
		\citenamefont {Gorodetsky},\ and\ \citenamefont
		{Kippenberg}}]{herr2014temporal}%
	\BibitemOpen
	\bibfield  {author} {\bibinfo {author} {\bibfnamefont {T.}~\bibnamefont
			{Herr}}, \bibinfo {author} {\bibfnamefont {V.}~\bibnamefont {Brasch}},
		\bibinfo {author} {\bibfnamefont {J.~D.}\ \bibnamefont {Jost}}, \bibinfo
		{author} {\bibfnamefont {C.~Y.}\ \bibnamefont {Wang}}, \bibinfo {author}
		{\bibfnamefont {N.~M.}\ \bibnamefont {Kondratiev}}, \bibinfo {author}
		{\bibfnamefont {M.~L.}\ \bibnamefont {Gorodetsky}}, \ and\ \bibinfo {author}
		{\bibfnamefont {T.~J.}\ \bibnamefont {Kippenberg}},\ }\href
	{https://www.nature.com/articles/nphoton.2013.343} {\bibfield  {journal}
		{\bibinfo  {journal} {Nat. Phot.}\ }\textbf {\bibinfo {volume} {8}},\
		\bibinfo {pages} {145} (\bibinfo {year} {2014}{\natexlab{a}})}\BibitemShut
	{NoStop}%
	\bibitem [{\citenamefont {H{\"o}nl}\ \emph {et~al.}(2018)\citenamefont
		{H{\"o}nl}, \citenamefont {Hahn}, \citenamefont {Baumgartner}, \citenamefont
		{Czornomaz},\ and\ \citenamefont {Seidler}}]{honl2018highly}%
	\BibitemOpen
	\bibfield  {author} {\bibinfo {author} {\bibfnamefont {S.}~\bibnamefont
			{H{\"o}nl}}, \bibinfo {author} {\bibfnamefont {H.}~\bibnamefont {Hahn}},
		\bibinfo {author} {\bibfnamefont {Y.}~\bibnamefont {Baumgartner}}, \bibinfo
		{author} {\bibfnamefont {L.}~\bibnamefont {Czornomaz}}, \ and\ \bibinfo
		{author} {\bibfnamefont {P.}~\bibnamefont {Seidler}},\ }\href
	{http://iopscience.iop.org/article/10.1088/1361-6463/aab8b7/meta} {\bibfield
		{journal} {\bibinfo  {journal} {J. Phys. D: Appl. Phys.}\ }\textbf {\bibinfo
			{volume} {51}},\ \bibinfo {pages} {185203} (\bibinfo {year}
		{2018})}\BibitemShut {NoStop}%
	\bibitem [{sup()}]{suppinfo}%
	\BibitemOpen
	\href@noop {} {\bibinfo  {journal} {See Supporting Information}\
	}\BibitemShut {NoStop}%
	\bibitem [{\citenamefont {Griffith}\ \emph {et~al.}(2015)\citenamefont
		{Griffith}, \citenamefont {Lau}, \citenamefont {Cardenas}, \citenamefont
		{Okawachi}, \citenamefont {Mohanty}, \citenamefont {Fain}, \citenamefont
		{Lee}, \citenamefont {Yu}, \citenamefont {Phare},\ and\ \citenamefont
		{Poitras}}]{griffith2015silicon}%
	\BibitemOpen
	\bibfield  {journal} {  }\bibfield  {author} {\bibinfo {author} {\bibfnamefont
			{A.~G.}\ \bibnamefont {Griffith}}, \bibinfo {author} {\bibfnamefont {R.~K.}\
			\bibnamefont {Lau}}, \bibinfo {author} {\bibfnamefont {J.}~\bibnamefont
			{Cardenas}}, \bibinfo {author} {\bibfnamefont {Y.}~\bibnamefont {Okawachi}},
		\bibinfo {author} {\bibfnamefont {A.}~\bibnamefont {Mohanty}}, \bibinfo
		{author} {\bibfnamefont {R.}~\bibnamefont {Fain}}, \bibinfo {author}
		{\bibfnamefont {Y.~H.~D.}\ \bibnamefont {Lee}}, \bibinfo {author}
		{\bibfnamefont {M.}~\bibnamefont {Yu}}, \bibinfo {author} {\bibfnamefont
			{C.~T.}\ \bibnamefont {Phare}}, \ and\ \bibinfo {author} {\bibfnamefont
			{C.~B.}\ \bibnamefont {Poitras}},\ }\href
	{https://www.nature.com/articles/ncomms7299} {\bibfield  {journal} {\bibinfo
			{journal} {Nat. Comm.}\ }\textbf {\bibinfo {volume} {6}},\ \bibinfo {pages}
		{6299} (\bibinfo {year} {2015})}\BibitemShut {NoStop}%
	\bibitem [{\citenamefont {Rokhsari}\ and\ \citenamefont
		{Vahala}(2005)}]{rokhsari2005observation}%
	\BibitemOpen
	\bibfield  {author} {\bibinfo {author} {\bibfnamefont {H.}~\bibnamefont
			{Rokhsari}}\ and\ \bibinfo {author} {\bibfnamefont {K.~J.}\ \bibnamefont
			{Vahala}},\ }\href
	{https://www.osapublishing.org/abstract.cfm?uri=OL-30-4-427} {\bibfield
		{journal} {\bibinfo  {journal} {Opt. Lett.}\ }\textbf {\bibinfo {volume}
			{30}},\ \bibinfo {pages} {427} (\bibinfo {year} {2005})}\BibitemShut
	{NoStop}%
	\bibitem [{\citenamefont {Kippenberg}\ \emph {et~al.}(2004)\citenamefont
		{Kippenberg}, \citenamefont {Spillane},\ and\ \citenamefont
		{Vahala}}]{kippenberg2004kerr}%
	\BibitemOpen
	\bibfield  {author} {\bibinfo {author} {\bibfnamefont {T.}~\bibnamefont
			{Kippenberg}}, \bibinfo {author} {\bibfnamefont {S.}~\bibnamefont
			{Spillane}}, \ and\ \bibinfo {author} {\bibfnamefont {K.}~\bibnamefont
			{Vahala}},\ }\href
	{https://journals.aps.org/prl/abstract/10.1103/PhysRevLett.93.083904}
	{\bibfield  {journal} {\bibinfo  {journal} {Physical review letters}\
		}\textbf {\bibinfo {volume} {93}},\ \bibinfo {pages} {083904} (\bibinfo
		{year} {2004})}\BibitemShut {NoStop}%
	\bibitem [{\citenamefont {Chembo}\ and\ \citenamefont
		{Yu}(2010)}]{chembo2010modal}%
	\BibitemOpen
	\bibfield  {author} {\bibinfo {author} {\bibfnamefont {Y.~K.}\ \bibnamefont
			{Chembo}}\ and\ \bibinfo {author} {\bibfnamefont {N.}~\bibnamefont {Yu}},\
	}\href {https://journals.aps.org/pra/abstract/10.1103/PhysRevA.82.033801}
	{\bibfield  {journal} {\bibinfo  {journal} {Physical Review A}\ }\textbf
		{\bibinfo {volume} {82}},\ \bibinfo {pages} {033801} (\bibinfo {year}
		{2010})}\BibitemShut {NoStop}%
	\bibitem [{\citenamefont {Xue}\ \emph {et~al.}(2017)\citenamefont {Xue},
		\citenamefont {Leo}, \citenamefont {Xuan}, \citenamefont
		{Jaramillo-Villegas}, \citenamefont {Wang}, \citenamefont {Leaird},
		\citenamefont {Erkintalo}, \citenamefont {Qi},\ and\ \citenamefont
		{Weiner}}]{xue2017second}%
	\BibitemOpen
	\bibfield  {author} {\bibinfo {author} {\bibfnamefont {X.}~\bibnamefont
			{Xue}}, \bibinfo {author} {\bibfnamefont {F.}~\bibnamefont {Leo}}, \bibinfo
		{author} {\bibfnamefont {Y.}~\bibnamefont {Xuan}}, \bibinfo {author}
		{\bibfnamefont {J.~A.}\ \bibnamefont {Jaramillo-Villegas}}, \bibinfo {author}
		{\bibfnamefont {P.-H.}\ \bibnamefont {Wang}}, \bibinfo {author}
		{\bibfnamefont {D.~E.}\ \bibnamefont {Leaird}}, \bibinfo {author}
		{\bibfnamefont {M.}~\bibnamefont {Erkintalo}}, \bibinfo {author}
		{\bibfnamefont {M.}~\bibnamefont {Qi}}, \ and\ \bibinfo {author}
		{\bibfnamefont {A.~M.}\ \bibnamefont {Weiner}},\ }\href {\doibase
		10.1038/lsa.2016.253} {\bibfield  {journal} {\bibinfo  {journal} {Light:
				Science \& Applications}\ }\textbf {\bibinfo {volume} {6}},\ \bibinfo {pages}
		{e16253} (\bibinfo {year} {2017})}\BibitemShut {NoStop}%
	\bibitem [{\citenamefont {Herr}(2013)}]{herr2013solitons}%
	\BibitemOpen
	\bibfield  {author} {\bibinfo {author} {\bibfnamefont {T.}~\bibnamefont
			{Herr}},\ }\href@noop {} {\bibinfo {title} {Solitons and dynamics of
			frequency comb formation in optical microresonators},\ } (\bibinfo {year}
	{2013})\BibitemShut {NoStop}%
	\bibitem [{\citenamefont {Herr}\ \emph {et~al.}(2012)\citenamefont {Herr},
		\citenamefont {Hartinger}, \citenamefont {Riemensberger}, \citenamefont
		{Wang}, \citenamefont {Gavartin}, \citenamefont {Holzwarth}, \citenamefont
		{Gorodetsky},\ and\ \citenamefont {Kippenberg}}]{herr2012universal}%
	\BibitemOpen
	\bibfield  {author} {\bibinfo {author} {\bibfnamefont {T.}~\bibnamefont
			{Herr}}, \bibinfo {author} {\bibfnamefont {K.}~\bibnamefont {Hartinger}},
		\bibinfo {author} {\bibfnamefont {J.}~\bibnamefont {Riemensberger}}, \bibinfo
		{author} {\bibfnamefont {C.}~\bibnamefont {Wang}}, \bibinfo {author}
		{\bibfnamefont {E.}~\bibnamefont {Gavartin}}, \bibinfo {author}
		{\bibfnamefont {R.}~\bibnamefont {Holzwarth}}, \bibinfo {author}
		{\bibfnamefont {M.~L.}\ \bibnamefont {Gorodetsky}}, \ and\ \bibinfo {author}
		{\bibfnamefont {T.}~\bibnamefont {Kippenberg}},\ }\href
	{https://www.nature.com/articles/nphoton.2012.127} {\bibfield  {journal}
		{\bibinfo  {journal} {Nat. Phot.}\ }\textbf {\bibinfo {volume} {6}},\
		\bibinfo {pages} {480} (\bibinfo {year} {2012})}\BibitemShut {NoStop}%
	\bibitem [{\citenamefont {Herr}\ \emph
		{et~al.}(2014{\natexlab{b}})\citenamefont {Herr}, \citenamefont {Brasch},
		\citenamefont {Jost}, \citenamefont {Mirgorodskiy}, \citenamefont {Lihachev},
		\citenamefont {Gorodetsky},\ and\ \citenamefont {Kippenberg}}]{herr2014mode}%
	\BibitemOpen
	\bibfield  {author} {\bibinfo {author} {\bibfnamefont {T.}~\bibnamefont
			{Herr}}, \bibinfo {author} {\bibfnamefont {V.}~\bibnamefont {Brasch}},
		\bibinfo {author} {\bibfnamefont {J.}~\bibnamefont {Jost}}, \bibinfo {author}
		{\bibfnamefont {I.}~\bibnamefont {Mirgorodskiy}}, \bibinfo {author}
		{\bibfnamefont {G.}~\bibnamefont {Lihachev}}, \bibinfo {author}
		{\bibfnamefont {M.}~\bibnamefont {Gorodetsky}}, \ and\ \bibinfo {author}
		{\bibfnamefont {T.}~\bibnamefont {Kippenberg}},\ }\href
	{https://journals.aps.org/prl/abstract/10.1103/PhysRevLett.113.123901}
	{\bibfield  {journal} {\bibinfo  {journal} {Phys. Rev. Lett.}\ }\textbf
		{\bibinfo {volume} {113}},\ \bibinfo {pages} {123901} (\bibinfo {year}
		{2014}{\natexlab{b}})}\BibitemShut {NoStop}%
	\bibitem [{\citenamefont {Guo}\ \emph {et~al.}(2016)\citenamefont {Guo},
		\citenamefont {Zou},\ and\ \citenamefont {Tang}}]{guo2016second}%
	\BibitemOpen
	\bibfield  {author} {\bibinfo {author} {\bibfnamefont {X.}~\bibnamefont
			{Guo}}, \bibinfo {author} {\bibfnamefont {C.-L.}\ \bibnamefont {Zou}}, \ and\
		\bibinfo {author} {\bibfnamefont {H.~X.}\ \bibnamefont {Tang}},\ }\href
	{https://www.osapublishing.org/abstract.cfm?uri=optica-3-10-1126} {\bibfield
		{journal} {\bibinfo  {journal} {Optica}\ }\textbf {\bibinfo {volume} {3}},\
		\bibinfo {pages} {1126} (\bibinfo {year} {2016})}\BibitemShut {NoStop}%
	\bibitem [{\citenamefont {Min}\ \emph {et~al.}(2005)\citenamefont {Min},
		\citenamefont {Yang},\ and\ \citenamefont {Vahala}}]{min2005controlled}%
	\BibitemOpen
	\bibfield  {author} {\bibinfo {author} {\bibfnamefont {B.}~\bibnamefont
			{Min}}, \bibinfo {author} {\bibfnamefont {L.}~\bibnamefont {Yang}}, \ and\
		\bibinfo {author} {\bibfnamefont {K.}~\bibnamefont {Vahala}},\ }\href
	{http://aip.scitation.org/doi/abs/10.1063/1.2120921} {\bibfield  {journal}
		{\bibinfo  {journal} {Appl. Phys. Lett.}\ }\textbf {\bibinfo {volume} {87}},\
		\bibinfo {pages} {181109} (\bibinfo {year} {2005})}\BibitemShut {NoStop}%
	\bibitem [{\citenamefont {Hausmann}\ \emph {et~al.}(2014)\citenamefont
		{Hausmann}, \citenamefont {Bulu}, \citenamefont {Venkataraman}, \citenamefont
		{Deotare},\ and\ \citenamefont {Lon{\v{c}}ar}}]{hausmann2014diamond}%
	\BibitemOpen
	\bibfield  {author} {\bibinfo {author} {\bibfnamefont {B.}~\bibnamefont
			{Hausmann}}, \bibinfo {author} {\bibfnamefont {I.}~\bibnamefont {Bulu}},
		\bibinfo {author} {\bibfnamefont {V.}~\bibnamefont {Venkataraman}}, \bibinfo
		{author} {\bibfnamefont {P.}~\bibnamefont {Deotare}}, \ and\ \bibinfo
		{author} {\bibfnamefont {M.}~\bibnamefont {Lon{\v{c}}ar}},\ }\href
	{https://www.nature.com/nphoton/journal/v8/n5/abs/nphoton.2014.72.html}
	{\bibfield  {journal} {\bibinfo  {journal} {Nat. Phot.}\ }\textbf {\bibinfo
			{volume} {8}},\ \bibinfo {pages} {369} (\bibinfo {year} {2014})}\BibitemShut
	{NoStop}%
	\bibitem [{\citenamefont {Jung}\ \emph {et~al.}(2013)\citenamefont {Jung},
		\citenamefont {Xiong}, \citenamefont {Fong}, \citenamefont {Zhang},\ and\
		\citenamefont {Tang}}]{jung2013optical}%
	\BibitemOpen
	\bibfield  {author} {\bibinfo {author} {\bibfnamefont {H.}~\bibnamefont
			{Jung}}, \bibinfo {author} {\bibfnamefont {C.}~\bibnamefont {Xiong}},
		\bibinfo {author} {\bibfnamefont {K.~Y.}\ \bibnamefont {Fong}}, \bibinfo
		{author} {\bibfnamefont {X.}~\bibnamefont {Zhang}}, \ and\ \bibinfo {author}
		{\bibfnamefont {H.~X.}\ \bibnamefont {Tang}},\ }\href
	{https://www.osapublishing.org/abstract.cfm?uri=ol-38-15-2810} {\bibfield
		{journal} {\bibinfo  {journal} {Opt. Lett.}\ }\textbf {\bibinfo {volume}
			{38}},\ \bibinfo {pages} {2810} (\bibinfo {year} {2013})}\BibitemShut
	{NoStop}%
	\bibitem [{\citenamefont {Ji}\ \emph {et~al.}(2017)\citenamefont {Ji},
		\citenamefont {Barbosa}, \citenamefont {Roberts}, \citenamefont {Dutt},
		\citenamefont {Cardenas}, \citenamefont {Okawachi}, \citenamefont {Bryant},
		\citenamefont {Gaeta},\ and\ \citenamefont {Lipson}}]{ji2017ultra}%
	\BibitemOpen
	\bibfield  {author} {\bibinfo {author} {\bibfnamefont {X.}~\bibnamefont
			{Ji}}, \bibinfo {author} {\bibfnamefont {F.~A.}\ \bibnamefont {Barbosa}},
		\bibinfo {author} {\bibfnamefont {S.~P.}\ \bibnamefont {Roberts}}, \bibinfo
		{author} {\bibfnamefont {A.}~\bibnamefont {Dutt}}, \bibinfo {author}
		{\bibfnamefont {J.}~\bibnamefont {Cardenas}}, \bibinfo {author}
		{\bibfnamefont {Y.}~\bibnamefont {Okawachi}}, \bibinfo {author}
		{\bibfnamefont {A.}~\bibnamefont {Bryant}}, \bibinfo {author} {\bibfnamefont
			{A.~L.}\ \bibnamefont {Gaeta}}, \ and\ \bibinfo {author} {\bibfnamefont
			{M.}~\bibnamefont {Lipson}},\ }\href
	{https://www.osapublishing.org/abstract.cfm?uri=optica-4-6-619} {\bibfield
		{journal} {\bibinfo  {journal} {Optica}\ }\textbf {\bibinfo {volume} {4}},\
		\bibinfo {pages} {619} (\bibinfo {year} {2017})}\BibitemShut {NoStop}%
	\bibitem [{\citenamefont {Razzari}\ \emph {et~al.}(2010)\citenamefont
		{Razzari}, \citenamefont {Duchesne}, \citenamefont {Ferrera}, \citenamefont
		{Morandotti}, \citenamefont {Chu}, \citenamefont {Little},\ and\
		\citenamefont {Moss}}]{razzari2010cmos}%
	\BibitemOpen
	\bibfield  {author} {\bibinfo {author} {\bibfnamefont {L.}~\bibnamefont
			{Razzari}}, \bibinfo {author} {\bibfnamefont {D.}~\bibnamefont {Duchesne}},
		\bibinfo {author} {\bibfnamefont {M.}~\bibnamefont {Ferrera}}, \bibinfo
		{author} {\bibfnamefont {R.}~\bibnamefont {Morandotti}}, \bibinfo {author}
		{\bibfnamefont {S.}~\bibnamefont {Chu}}, \bibinfo {author} {\bibfnamefont
			{B.}~\bibnamefont {Little}}, \ and\ \bibinfo {author} {\bibfnamefont
			{D.}~\bibnamefont {Moss}},\ }\href
	{https://www.nature.com/nphoton/journal/v4/n1/abs/nphoton.2009.236.html}
	{\bibfield  {journal} {\bibinfo  {journal} {Nat. Phot.}\ }\textbf {\bibinfo
			{volume} {4}},\ \bibinfo {pages} {41} (\bibinfo {year} {2010})}\BibitemShut
	{NoStop}%
\end{thebibliography}

\begin{thebibliography}{14}%
	\makeatletter
	\providecommand \@ifxundefined [1]{%
		\@ifx{#1\undefined}
	}%
	\providecommand \@ifnum [1]{%
		\ifnum #1\expandafter \@firstoftwo
		\else \expandafter \@secondoftwo
		\fi
	}%
	\providecommand \@ifx [1]{%
		\ifx #1\expandafter \@firstoftwo
		\else \expandafter \@secondoftwo
		\fi
	}%
	\providecommand \natexlab [1]{#1}%
	\providecommand \enquote  [1]{``#1''}%
	\providecommand \bibnamefont  [1]{#1}%
	\providecommand \bibfnamefont [1]{#1}%
	\providecommand \citenamefont [1]{#1}%
	\providecommand \href@noop [0]{\@secondoftwo}%
	\providecommand \href [0]{\begingroup \@sanitize@url \@href}%
	\providecommand \@href[1]{\@@startlink{#1}\@@href}%
	\providecommand \@@href[1]{\endgroup#1\@@endlink}%
	\providecommand \@sanitize@url [0]{\catcode `\\12\catcode `\$12\catcode
		`\&12\catcode `\#12\catcode `\^12\catcode `\_12\catcode `\%12\relax}%
	\providecommand \@@startlink[1]{}%
	\providecommand \@@endlink[0]{}%
	\providecommand \url  [0]{\begingroup\@sanitize@url \@url }%
	\providecommand \@url [1]{\endgroup\@href {#1}{\urlprefix }}%
	\providecommand \urlprefix  [0]{URL }%
	\providecommand \Eprint [0]{\href }%
	\providecommand \doibase [0]{http://dx.doi.org/}%
	\providecommand \selectlanguage [0]{\@gobble}%
	\providecommand \bibinfo  [0]{\@secondoftwo}%
	\providecommand \bibfield  [0]{\@secondoftwo}%
	\providecommand \translation [1]{[#1]}%
	\providecommand \BibitemOpen [0]{}%
	\providecommand \bibitemStop [0]{}%
	\providecommand \bibitemNoStop [0]{.\EOS\space}%
	\providecommand \EOS [0]{\spacefactor3000\relax}%
	\providecommand \BibitemShut  [1]{\csname bibitem#1\endcsname}%
	\let\auto@bib@innerbib\@empty
	\bibitem [{\citenamefont {Agrawal}(2000)}]{agrawal2000nonlinear_B}%
	\BibitemOpen
	\bibfield  {author} {\bibinfo {author} {\bibfnamefont {G.~P.}\ \bibnamefont
			{Agrawal}},\ }\href
	{https://link.springer.com/chapter/10.1007/3-540-46629-0_9} {\emph {\bibinfo
			{title} {Nonlinear fiber optics}}}\ (\bibinfo  {publisher} {Springer},\
	\bibinfo {year} {2000})\BibitemShut {NoStop}%
	\bibitem [{\citenamefont {Wei}\ \emph {et~al.}(2018)\citenamefont {Wei},
		\citenamefont {Murray}, \citenamefont {Barnes}, \citenamefont {Krein},
		\citenamefont {Schunemann},\ and\ \citenamefont
		{Guha}}]{wei2018temperature_B}%
	\BibitemOpen
	\bibfield  {author} {\bibinfo {author} {\bibfnamefont {J.}~\bibnamefont
			{Wei}}, \bibinfo {author} {\bibfnamefont {J.~M.}\ \bibnamefont {Murray}},
		\bibinfo {author} {\bibfnamefont {J.~O.}\ \bibnamefont {Barnes}}, \bibinfo
		{author} {\bibfnamefont {D.~M.}\ \bibnamefont {Krein}}, \bibinfo {author}
		{\bibfnamefont {P.~G.}\ \bibnamefont {Schunemann}}, \ and\ \bibinfo {author}
		{\bibfnamefont {S.}~\bibnamefont {Guha}},\ }\href
	{https://www.osapublishing.org/abstract.cfm?uri=ome-8-2-485} {\bibfield
		{journal} {\bibinfo  {journal} {Optical Materials Express}\ }\textbf
		{\bibinfo {volume} {8}},\ \bibinfo {pages} {485} (\bibinfo {year}
		{2018})}\BibitemShut {NoStop}%
	\bibitem [{\citenamefont {Matsko}\ \emph {et~al.}(2005)\citenamefont {Matsko},
		\citenamefont {Savchenkov},\ and\ \citenamefont
		{Maleki}}]{matsko2005optical_B}%
	\BibitemOpen
	\bibfield  {author} {\bibinfo {author} {\bibfnamefont {A.}~\bibnamefont
			{Matsko}}, \bibinfo {author} {\bibfnamefont {A.}~\bibnamefont {Savchenkov}},
		\ and\ \bibinfo {author} {\bibfnamefont {L.}~\bibnamefont {Maleki}},\ }\href
	{https://link.aps.org/doi/10.1103/PhysRevA.71.033804} {\bibfield  {journal}
		{\bibinfo  {journal} {Phys. Rev. A}\ }\textbf {\bibinfo {volume} {71}},\
		\bibinfo {pages} {033804} (\bibinfo {year} {2005})}\BibitemShut {NoStop}%
	\bibitem [{\citenamefont {Chembo}\ and\ \citenamefont
		{Yu}(2010)}]{chembo2010modal_B}%
	\BibitemOpen
	\bibfield  {author} {\bibinfo {author} {\bibfnamefont {Y.~K.}\ \bibnamefont
			{Chembo}}\ and\ \bibinfo {author} {\bibfnamefont {N.}~\bibnamefont {Yu}},\
	}\href {https://journals.aps.org/pra/abstract/10.1103/PhysRevA.82.033801}
	{\bibfield  {journal} {\bibinfo  {journal} {Phys. Rev. A}\ }\textbf {\bibinfo
			{volume} {82}},\ \bibinfo {pages} {033801} (\bibinfo {year}
		{2010})}\BibitemShut {NoStop}%
	\bibitem [{\citenamefont {Herr}\ \emph {et~al.}(2014)\citenamefont {Herr},
		\citenamefont {Brasch}, \citenamefont {Jost}, \citenamefont {Wang},
		\citenamefont {Kondratiev}, \citenamefont {Gorodetsky},\ and\ \citenamefont
		{Kippenberg}}]{herr2014temporal_B}%
	\BibitemOpen
	\bibfield  {author} {\bibinfo {author} {\bibfnamefont {T.}~\bibnamefont
			{Herr}}, \bibinfo {author} {\bibfnamefont {V.}~\bibnamefont {Brasch}},
		\bibinfo {author} {\bibfnamefont {J.~D.}\ \bibnamefont {Jost}}, \bibinfo
		{author} {\bibfnamefont {C.~Y.}\ \bibnamefont {Wang}}, \bibinfo {author}
		{\bibfnamefont {N.~M.}\ \bibnamefont {Kondratiev}}, \bibinfo {author}
		{\bibfnamefont {M.~L.}\ \bibnamefont {Gorodetsky}}, \ and\ \bibinfo {author}
		{\bibfnamefont {T.~J.}\ \bibnamefont {Kippenberg}},\ }\href
	{https://www.nature.com/articles/nphoton.2013.343} {\bibfield  {journal}
		{\bibinfo  {journal} {Nat. Phot.}\ }\textbf {\bibinfo {volume} {8}},\
		\bibinfo {pages} {145} (\bibinfo {year} {2014})}\BibitemShut {NoStop}%
	\bibitem [{\citenamefont {Herr}(2013)}]{herr2013solitons_B}%
	\BibitemOpen
	\bibfield  {author} {\bibinfo {author} {\bibfnamefont {T.}~\bibnamefont
			{Herr}},\ }\href {https://infoscience.epfl.ch/record/188349} {\bibinfo
		{title} {Solitons and dynamics of frequency comb formation in optical
			microresonators},\ } (\bibinfo {year} {2013})\BibitemShut {NoStop}%
	\bibitem [{\citenamefont {Pu}\ \emph {et~al.}(2016)\citenamefont {Pu},
		\citenamefont {Ottaviano}, \citenamefont {Semenova},\ and\ \citenamefont
		{Yvind}}]{pu2016efficient_B}%
	\BibitemOpen
	\bibfield  {author} {\bibinfo {author} {\bibfnamefont {M.}~\bibnamefont
			{Pu}}, \bibinfo {author} {\bibfnamefont {L.}~\bibnamefont {Ottaviano}},
		\bibinfo {author} {\bibfnamefont {E.}~\bibnamefont {Semenova}}, \ and\
		\bibinfo {author} {\bibfnamefont {K.}~\bibnamefont {Yvind}},\ }\href
	{https://www.osapublishing.org/abstract.cfm?uri=optica-3-8-823} {\bibfield
		{journal} {\bibinfo  {journal} {Optica}\ }\textbf {\bibinfo {volume} {3}},\
		\bibinfo {pages} {823} (\bibinfo {year} {2016})}\BibitemShut {NoStop}%
	\bibitem [{\citenamefont {Hausmann}\ \emph {et~al.}(2014)\citenamefont
		{Hausmann}, \citenamefont {Bulu}, \citenamefont {Venkataraman}, \citenamefont
		{Deotare},\ and\ \citenamefont {Lon{\v{c}}ar}}]{hausmann2014diamond_B}%
	\BibitemOpen
	\bibfield  {author} {\bibinfo {author} {\bibfnamefont {B.}~\bibnamefont
			{Hausmann}}, \bibinfo {author} {\bibfnamefont {I.}~\bibnamefont {Bulu}},
		\bibinfo {author} {\bibfnamefont {V.}~\bibnamefont {Venkataraman}}, \bibinfo
		{author} {\bibfnamefont {P.}~\bibnamefont {Deotare}}, \ and\ \bibinfo
		{author} {\bibfnamefont {M.}~\bibnamefont {Lon{\v{c}}ar}},\ }\href
	{https://www.nature.com/nphoton/journal/v8/n5/abs/nphoton.2014.72.html}
	{\bibfield  {journal} {\bibinfo  {journal} {Nat. Phot.}\ }\textbf {\bibinfo
			{volume} {8}},\ \bibinfo {pages} {369} (\bibinfo {year} {2014})}\BibitemShut
	{NoStop}%
	\bibitem [{\citenamefont {Brasch}\ \emph {et~al.}(2016)\citenamefont {Brasch},
		\citenamefont {Geiselmann}, \citenamefont {Herr}, \citenamefont {Lihachev},
		\citenamefont {Pfeiffer}, \citenamefont {Gorodetsky},\ and\ \citenamefont
		{Kippenberg}}]{brasch2016photonic_B}%
	\BibitemOpen
	\bibfield  {author} {\bibinfo {author} {\bibfnamefont {V.}~\bibnamefont
			{Brasch}}, \bibinfo {author} {\bibfnamefont {M.}~\bibnamefont {Geiselmann}},
		\bibinfo {author} {\bibfnamefont {T.}~\bibnamefont {Herr}}, \bibinfo {author}
		{\bibfnamefont {G.}~\bibnamefont {Lihachev}}, \bibinfo {author}
		{\bibfnamefont {M.~H.}\ \bibnamefont {Pfeiffer}}, \bibinfo {author}
		{\bibfnamefont {M.~L.}\ \bibnamefont {Gorodetsky}}, \ and\ \bibinfo {author}
		{\bibfnamefont {T.~J.}\ \bibnamefont {Kippenberg}},\ }\href
	{http://science.sciencemag.org/content/351/6271/357.short} {\bibfield
		{journal} {\bibinfo  {journal} {Science}\ }\textbf {\bibinfo {volume}
			{351}},\ \bibinfo {pages} {357} (\bibinfo {year} {2016})}\BibitemShut
	{NoStop}%
	\bibitem [{\citenamefont {Rokhsari}\ and\ \citenamefont
		{Vahala}(2005)}]{rokhsari2005observation_B}%
	\BibitemOpen
	\bibfield  {author} {\bibinfo {author} {\bibfnamefont {H.}~\bibnamefont
			{Rokhsari}}\ and\ \bibinfo {author} {\bibfnamefont {K.~J.}\ \bibnamefont
			{Vahala}},\ }\href
	{https://www.osapublishing.org/abstract.cfm?uri=OL-30-4-427} {\bibfield
		{journal} {\bibinfo  {journal} {Opt. Lett.}\ }\textbf {\bibinfo {volume}
			{30}},\ \bibinfo {pages} {427} (\bibinfo {year} {2005})}\BibitemShut
	{NoStop}%
	\bibitem [{\citenamefont {Gorodetksy}\ \emph {et~al.}(2010)\citenamefont
		{Gorodetksy}, \citenamefont {Schliesser}, \citenamefont {Anetsberger},
		\citenamefont {Deleglise},\ and\ \citenamefont
		{Kippenberg}}]{gorodetksy2010determination_B}%
	\BibitemOpen
	\bibfield  {author} {\bibinfo {author} {\bibfnamefont {M.}~\bibnamefont
			{Gorodetksy}}, \bibinfo {author} {\bibfnamefont {A.}~\bibnamefont
			{Schliesser}}, \bibinfo {author} {\bibfnamefont {G.}~\bibnamefont
			{Anetsberger}}, \bibinfo {author} {\bibfnamefont {S.}~\bibnamefont
			{Deleglise}}, \ and\ \bibinfo {author} {\bibfnamefont {T.~J.}\ \bibnamefont
			{Kippenberg}},\ }\href
	{https://www.osapublishing.org/abstract.cfm?uri=oe-18-22-23236} {\bibfield
		{journal} {\bibinfo  {journal} {Optics express}\ }\textbf {\bibinfo {volume}
			{18}},\ \bibinfo {pages} {23236} (\bibinfo {year} {2010})}\BibitemShut
	{NoStop}%
	\bibitem [{\citenamefont {Herr}\ \emph {et~al.}(2012)\citenamefont {Herr},
		\citenamefont {Hartinger}, \citenamefont {Riemensberger}, \citenamefont
		{Wang}, \citenamefont {Gavartin}, \citenamefont {Holzwarth}, \citenamefont
		{Gorodetsky},\ and\ \citenamefont {Kippenberg}}]{herr2012universal_B}%
	\BibitemOpen
	\bibfield  {author} {\bibinfo {author} {\bibfnamefont {T.}~\bibnamefont
			{Herr}}, \bibinfo {author} {\bibfnamefont {K.}~\bibnamefont {Hartinger}},
		\bibinfo {author} {\bibfnamefont {J.}~\bibnamefont {Riemensberger}}, \bibinfo
		{author} {\bibfnamefont {C.}~\bibnamefont {Wang}}, \bibinfo {author}
		{\bibfnamefont {E.}~\bibnamefont {Gavartin}}, \bibinfo {author}
		{\bibfnamefont {R.}~\bibnamefont {Holzwarth}}, \bibinfo {author}
		{\bibfnamefont {M.~L.}\ \bibnamefont {Gorodetsky}}, \ and\ \bibinfo {author}
		{\bibfnamefont {T.}~\bibnamefont {Kippenberg}},\ }\href
	{https://www.nature.com/articles/nphoton.2012.127} {\bibfield  {journal}
		{\bibinfo  {journal} {Nat. Phot.}\ }\textbf {\bibinfo {volume} {6}},\
		\bibinfo {pages} {480} (\bibinfo {year} {2012})}\BibitemShut {NoStop}%
	\bibitem [{\citenamefont {Schneider}\ \emph {et~al.}(2018)\citenamefont
		{Schneider}, \citenamefont {Welter}, \citenamefont {Baumgartner},
		\citenamefont {Hahn}, \citenamefont {Czornomaz},\ and\ \citenamefont
		{Seidler}}]{schneider2018gallium_B}%
	\BibitemOpen
	\bibfield  {author} {\bibinfo {author} {\bibfnamefont {K.}~\bibnamefont
			{Schneider}}, \bibinfo {author} {\bibfnamefont {P.}~\bibnamefont {Welter}},
		\bibinfo {author} {\bibfnamefont {Y.}~\bibnamefont {Baumgartner}}, \bibinfo
		{author} {\bibfnamefont {H.}~\bibnamefont {Hahn}}, \bibinfo {author}
		{\bibfnamefont {L.}~\bibnamefont {Czornomaz}}, \ and\ \bibinfo {author}
		{\bibfnamefont {P.}~\bibnamefont {Seidler}},\ }\href
	{https://www.osapublishing.org/abstract.cfm?uri=jlt-36-14-2994} {\bibfield
		{journal} {\bibinfo  {journal} {Journal of Lightwave Technology}\ }\textbf
		{\bibinfo {volume} {36}},\ \bibinfo {pages} {2994} (\bibinfo {year}
		{2018})}\BibitemShut {NoStop}%
	\bibitem [{\citenamefont {H{\"o}nl}\ \emph {et~al.}(2018)\citenamefont
		{H{\"o}nl}, \citenamefont {Hahn}, \citenamefont {Baumgartner}, \citenamefont
		{Czornomaz},\ and\ \citenamefont {Seidler}}]{honl2018highly_B}%
	\BibitemOpen
	\bibfield  {author} {\bibinfo {author} {\bibfnamefont {S.}~\bibnamefont
			{H{\"o}nl}}, \bibinfo {author} {\bibfnamefont {H.}~\bibnamefont {Hahn}},
		\bibinfo {author} {\bibfnamefont {Y.}~\bibnamefont {Baumgartner}}, \bibinfo
		{author} {\bibfnamefont {L.}~\bibnamefont {Czornomaz}}, \ and\ \bibinfo
		{author} {\bibfnamefont {P.}~\bibnamefont {Seidler}},\ }\href
	{http://iopscience.iop.org/article/10.1088/1361-6463/aab8b7/meta} {\bibfield
		{journal} {\bibinfo  {journal} {J. Phys. D: Appl. Phys.}\ }\textbf {\bibinfo
			{volume} {51}},\ \bibinfo {pages} {185203} (\bibinfo {year}
		{2018})}\BibitemShut {NoStop}%
\end{thebibliography}
\end{document}